\documentclass{llncs}

\usepackage[usenames,dvipsnames]{color}

\usepackage{amsmath} 
\usepackage{amsxtra} 
\usepackage{amssymb}
\usepackage{mathabx}
\usepackage{mathtools}
\usepackage{textcomp}

\usepackage{latexsym}
\usepackage{pslatex}
\usepackage{epsfig}
\usepackage{wrapfig}
\usepackage{enumerate}
\usepackage{cancel}

\usepackage{paralist}

\usepackage{ listings }

\usepackage{algorithm}
\usepackage{algorithmicx}
\usepackage[noend]{algpseudocode}

\spnewtheorem{ass}[theorem]{Assumption}{\bfseries}{\itshape}
\spnewtheorem{fact}[theorem]{Fact}{\bfseries}{\itshape}

\renewcommand{\iff}{\Leftrightarrow}

\newcommand{\topleft}[1] {{~}^{\scriptscriptstyle{\blacksquare}} \! M}
\newcommand{\botleft}[1] {{~}_{\scriptscriptstyle{\blacksquare}} \! M}
\newcommand{\topright}[1] {M^{\scriptscriptstyle{\blacksquare}}}
\newcommand{\botright}[1] {M_{\scriptscriptstyle{\blacksquare}}}

\newcommand{\rbr}{{\bf ]\!]}}
\newcommand{\lbr}{{\bf [\![}}
\newcommand{\sem}[1]{\lbr #1 \rbr}

\newcommand{\thd}{\mbox{Th}(\mathcal{D})}
\newcommand{\smallthd}{\mbox{\tiny Th}(\mathcal{D})}
\newcommand{\modelsthd}{\models_{\mbox{\tiny Th}(\scriptscriptstyle\mathcal{D})}}

\renewcommand{\vec}[1]{{\bf {#1}}}

\newcommand{\true}{\top}
\newcommand{\false}{\bot}

\newcommand{\lang}[1]{{\mathcal L}({#1})}

\newcommand{\arrow}[1]{\xrightarrow{{\scriptscriptstyle #1}}}

\newcommand{\nat}{{\bf \mathbb{N}}}

\def\vr{\kern-\arraycolsep & \kern-\arraycolsep}
\def\VR{\kern-\arraycolsep\strut\vrule}

\def\set#1{{\left\{ #1 \right\}}}

\def\tuple#1{{\langle #1 \rangle}}

\def\len#1{{|{#1}|}}
\def\prod{\Delta}

\DeclareFontFamily{U}{mathx}{\hyphenchar\font45}
\DeclareFontShape{U}{mathx}{m}{n}{
      <5> <6> <7> <8> <9> <10>
      <10.95> <12> <14.4> <17.28> <20.74> <24.88>
      mathx10
      }{}
\DeclareSymbolFont{mathx}{U}{mathx}{m}{n}
\DeclareFontSubstitution{U}{mathx}{m}{n}
\DeclareMathAccent{\widecheck}{0}{mathx}{"71}
\DeclareMathAccent{\wideparen}{0}{mathx}{"75}

\renewcommand{\vec}[1]{{\mathbf {#1}}}

\def\proj{\mathbin{\downarrow}}

\newcommand{\old}{{\mathit{old}}}
\newcommand{\new}{{\mathit{new}}}

\definecolor{darkgreen}{rgb}{0,0.6,0}

\begin{document}

\title{Abstraction Refinement and Antichains for Trace Inclusion of Infinite
State Systems}

\author{Radu Iosif \and Adam Rogalewicz \and Tom\'{a}\v{s}~Vojnar}
  
\institute{CNRS/Verimag, France and FIT BUT, Czech Republic}

\maketitle

\begin{abstract}A \emph{data automaton} is a finite automaton
equipped with variables (counters or registers) ranging over infinite data
domains. A trace of a data automaton is an alternating sequence of alphabet
symbols and values taken by the counters during an execution of the automaton.
The problem addressed in this paper is the inclusion between the sets of traces
(data languages) recognized by such automata. Since the problem is undecidable
in general, we give a semi-algorithm based on abstraction refinement, which is
proved to be sound and complete, but whose termination is not guaranteed. We
have implemented our technique in a~prototype tool and show promising results on
several non-trivial examples.\end{abstract}

\section{Introduction}

In this paper, we address a \emph{trace inclusion} problem for
infinite-state systems. Given \begin{inparaenum}[(i)]
  \item a network of \emph{data automata}
    $\mathcal{A}=\tuple{A_1,\ldots,A_N}$ that communicate via a set of
    shared variables $\vec{x}_{\mathcal{A}}$, ranging over an infinite
    data domain, and a set of input events $\Sigma_{\mathcal{A}}$, and
  \item a~data automaton $B$ whose set of variables $\vec{x}_B$ is a
    subset of $\vec{x}_{\mathcal{A}}$,
\end{inparaenum} does the set of (finite) traces of $B$ contain the traces of
$A$? Here, by a \emph{trace}, we understand an alternating sequence of
valuations of the variables from the set $\vec{x}_B$ and input events
from the set $\Sigma_{\mathcal{A}} \cap \Sigma_B$, starting and ending
with a valuation. 
Typically, the network of automata $\mathcal{A}$ is an implementation
of a concurrent system and $B$ is a specification of the set of good
behaviors of the system.

\begin{figure}[t]
\begin{center}
  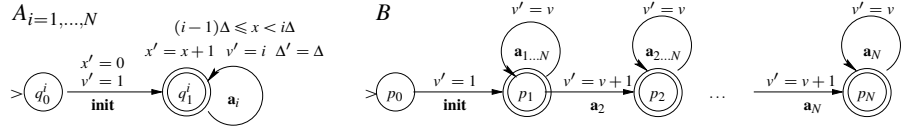
  \caption{An instance of the trace inclusion problem.}
  \label{fig:running-example}
\end{center}
\end{figure}

Consider, for instance, the network $\tuple{A_1, \ldots, A_N}$ of data
automata equipped with the integer-valued variables $x$ and $v$ shown
in Fig.~ \ref{fig:running-example}--left. The automata synchronize on
the $\mathbf{init}$ symbol and interleave their
$\mathbf{a}_{1,\ldots,N}$ actions.  Each automaton $A_i$ increases the
shared variable $x$ and writes its identifier $i$ into the shared
variable $v$ as long as the value of $x$ is in the interval
$[(i-1)\Delta,i\Delta-1]$, and it is inactive outside this interval,
where $\Delta\geq1$ is an unbounded parameter of the network. A
possible specification for this network might require that each firing
sequence is of the form $\mathbf{init}~ \mathbf{a}_{1,\ldots,N}^*
~\mathbf{a}_2~ \mathbf{a}_{2,\ldots,N}^* \ldots \mathbf{a}_i~
\mathbf{a}_i^*$ for some $1 \leq i \leq N$, and that $v$ is increased
only on the first occurrence of the events
$\mathbf{a}_2,\ldots,\mathbf{a}_i$, in this order. This condition is
encoded by the automaton $B$
(Fig. \ref{fig:running-example}--right). Observe that only the $v$
variable is shared between the network $\tuple{A_1,\ldots,A_N}$ and
the specification automaton $B$---we say that $v$ is \emph{observable}
in this case.  An example of a trace, for $\Delta=2$ and $N\geq3$, is:
\((v=0) ~\mathbf{init}~ (v=1) ~\mathbf{a}_1~ (v=1) ~\mathbf{a}_1~
(v=1)$ $\mathbf{a}_2~ (v=2) ~\mathbf{a}_2~ (v=2) ~\mathbf{a}_3~
(v=3)\). Our problem is to check that this, and all other traces of
the network, are included in the language of the specification
automaton, called the \emph{observer}. The trace inclusion problem
has multiple applications,
e.g.:\begin{itemize}\setlength{\itemsep}{1mm}

    \item Decision procedures for logics describing array structures
      within imperative programs \cite{lia,lpar08} that use a
      translation of array formulae to integer counter automata which
      encode the set of array models of a formula. The expressiveness
      of such logics is currently limited by the decidability of the
      emptiness (reachability) problem for counter automata. If we
      give up on decidability, we can reduce an entailment between two
      array formulae to the trace inclusion of two integer counter
      automata, and use the method presented in this paper as a
      semi-decision procedure. To assess this claim, we have applied
      our trace inclusion method to several verification conditions
      for programs with unbounded arrays of integers \cite{cav09}.

    \item Timed automata and regular specifications of timed languages
      \cite{AlurDill94} can be both represented by finite automata
      extended with real-valued variables
      \cite{fribourg:timedAutomata}. The verification problem boils
      down to the trace inclusion of two real-valued data automata.
      In this context, our method has been tested on several timed
      verification problems, including communication protocols and
      boolean circuits \cite{stavros-thesis}.
\end{itemize} 

When developing a method for checking the inclusion between trace
languages of automata extended with variables ranging over infinite
data domains, the first problem is the lack of determinisation and/or
complementation results. In fact, certain classes of infinite state
systems, such as timed automata \cite{AlurDill94}, cannot be
determinized and are provably not closed under complement. This is the
case due to the fact that the clock variables of a timed automaton are
not observable in its timed language, which records only the time
lapses between successive events. However, if we require that the
values of all variables of a data automaton be part of its trace
language, we obtain a determinisation result, which generalizes the
classical subset construction by taking into account the data
valuations. Building on this first result, we define the complement of
a data language and reduce the trace inclusion problem to the
emptiness of a product data automaton $\mathcal{L}(A \times
\overline{B}) = \emptyset$. It is crucial, for this reduction, that
the variables $\vec{x}_B$ of the right-hand side data automaton $B$
(the one being determinized) are also controlled by the left-hand side
automaton $A$, in other words, that $B$ has no hidden variables.

The language emptiness problem for data automata is, in general,
undecidable \cite{minsky67}. Nevertheless, several semi-algorithms and
tools for this problem (better known as the \emph{reachability}
problem) have been developed
\cite{fast,blast,mcmillan06,rybal-pldi11}. Among those, the technique
of \emph{lazy predicate abstraction} \cite{blast} combined with
\emph{counterexample-driven refinement} using \emph{interpolants}
\cite{mcmillan06} has been shown to be particularly successful in
proving emptiness of rather large infinite-state systems. Moreover,
this technique shares similar aspects with the antichain-based
algorithm for language inclusion in the case of a finite alphabet
\cite{abdulla}. An important similarity is that both techniques use a
partial order on states, to prune the state space during the search.

The main result of this paper is a semi-algorithm that combines the
principle of the antichain-based language inclusion algorithm
\cite{abdulla} with the interpolant-based abstraction refinement
semi-algorithm \cite{mcmillan06}, via a general notion of
language-based subsumption relation. We have implemented our
semi-algorithm in a prototype tool and carried out a~number of
experiments, involving hardware, real-time systems, and array logic
problems. Since our procedure tests inclusion within a set of good
traces, instead of empty intersection with a set of error traces, we
can encode rather complex verification conditions concisely, using
automata of relatively small size.

\subsection{Overview}

\begin{figure}[htb]
\begin{center}
  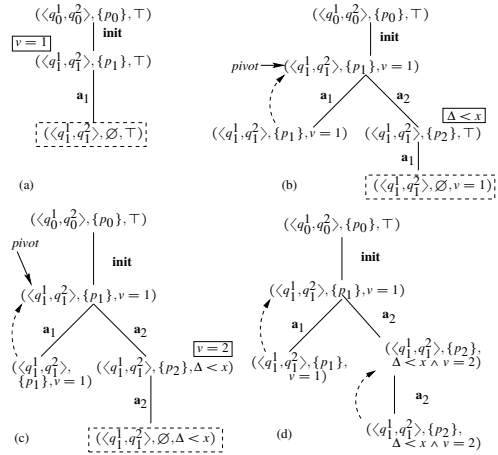
  \caption{Sample run of our semi-algorithm.}
  \label{fig:refinement}
\end{center}
\end{figure}

We introduce the reader to our trace inclusion method by means of an
example. Let us consider the network of data automata
$\tuple{A_1,A_2}$ and the data automaton $B$ from
Fig.~\ref{fig:running-example}. We prove that, for any value of
$\Delta$, any trace of the network $\tuple{A_1,A_2}$, obtained as an
interleaving of the actions of $A_1$ and $A_2$, is also a~trace of the
observer $B$. To this end, our procedure will fire increasingly longer
sequences of input events, in search for a counterexample trace. We
keep a set of predicates associated with each state
$(\tuple{q_1,q_2},P)$ of the product automaton where $q_i$ is a state
of $A_i$ and $P$ is a~set of states of $B$. These predicates are
formulae that define over-approximations of the data values reached
simultaneously by the network, when $A_i$ is the state $q_i$, and by
the observer $B$, in every state from $P$.

The first input event is $\mathbf{init}$, on which $A_1$ and $A_2$
synchronize, moving together from the initial state $\tuple{q_0^1,
  q_0^2}$ to $\tuple{q_1^1,q_1^2}$. In response, $B$ can chose to
either\begin{inparaenum}[(i)]
\item move from $\set{p_0}$ to $\set{p_1}$, matching the only
  transition rule from $p_0$, or
\item ignore the transition rule and move to the empty set. 
\end{inparaenum}
In the first case, the values of $v$ match the relation of the rule
$p_0 \arrow{\mathbf{init},v'=1}{} p_1$, while in the second case, these
values match the negated relation $\neg(v'=1)$. The second case is
impossible because the action of the network requires $x'=0 \wedge
v'=1$. The only successor state is thus
$(\tuple{q_1^1,q_1^2},\set{p_1})$ in Fig. \ref{fig:refinement}
(a). Since no predicates are initially available at this state, the
best over-approximation of the set of reachable data valuations is the
universal set ($\true$).

The second input event is $\mathbf{a}_1$, on which $A_1$ moves from
$q_1^1$ back to itself, while $A_2$ makes an idle step because no
transition with $\mathbf{a}_1$ is enabled from $q^2_1$. Again, $B$ has
the choice between moving from $\set{p_1}$ either to $\emptyset$ or
$\set{p_1}$. Let us consider the first case, in which the successor
state is $(\tuple{q_1^1,q_1^2}, \emptyset, \true)$. Since $q_1^1$ and
$q^2_1$ are final states of $A_1$ and $A_2$, respectively, and no
final state of $B$ is present in $\emptyset$, we say that the state is
accepting. If the accepting state (in dashed boxes in
Fig. \ref{fig:refinement}) is reachable according to the transition
constraints along the input sequence $\mathbf{init}.\mathbf{a}_1$, we
have found a counterexample trace that is in the language of
$\tuple{A_1,A_2}$ but not in the language of $B$.

\enlargethispage{4mm}

To verify the reachability of the accepting state, we check the
satisfiability of the path formula corresponding to the composition of
the transition constraints $x'=0\wedge v'=1$ ($\mathbf{init}$) and $0
\leq x < \Delta \wedge x'=x+1 \wedge v'=1 \wedge \neg(v'=v)$
($\mathbf{a}_1$) in Fig. \ref{fig:refinement} (a). This formula is
unsatisfiable, and the proof of infeasibility provides the interpolant
$\tuple{v=1}$. This formula is an explanation for the infeasibility of
the path because it is implied by the first constraint and it is
unsatisfiable in conjunction with the second constraint. By
associating the new predicate $v=1$ with the state
$(\tuple{q_1^1,q_1^2}, \set{p_1})$, we ensure that the same spurious
path will never be explored again.

We delete the spurious counterexample and recompute the states along
the input sequence $\mathbf{init}.\mathbf{a}_1$ with the new
predicate. In this case, $(\tuple{q_1^1,q_1^2}, \emptyset)$ is
unreachable, and the outcome is $(\tuple{q_1^1,q_1^2},
\set{p_1},v=1)$. However, this state was first encountered after the
sequence $\mathbf{init}$, so there is no need to store a second
occurrence of this state in the tree. We say that the node
$\mathbf{init}.\mathbf{a}_1$ is subsumed by $\mathbf{init}$, and
indicate this by a dashed arrow in Fig. \ref{fig:refinement} (b).

We continue with $\mathbf{a}_2$ from the state $(\tuple{q_1^1,q_1^2},
\set{p_1},v=1)$. In this case, $A_1$ makes an idle step and $A_2$
moves from $q_1^2$ to itself. In response, $B$ has the choice between
moving from $\set{p_1}$ to either\begin{inparaenum}[(i)]
\item $\set{p_1}$ with the constraint $v'=v$, 
\item $\set{p_2}$ with the constraint $v'=v+1$, 
\item $\set{p_1,p_2}$ with the constraint $v'=v \wedge v'=v+1 \rightarrow
  \false$ (this possibility is discarded),
\item $\emptyset$ for  data values that satisfy $\neg(v'=v) \wedge \neg(v'=v+1)$.
\end{inparaenum} 
The last case is also discarded because the value of $v$ after
$\mathbf{init}$ constrained to $1$ and the $A_2$ imposes further the
constraint $v'=2$ and $v=1 \wedge v'=2 \wedge \neg(v'=v) \wedge
\neg(v'=v+1) \rightarrow \false$. Hence, the only
$\mathbf{a}_2$-successor of $(\tuple{q_1^1,q_1^2}, \set{p_1},v=1)$ is
$(\tuple{q_1^1,q_1^2}, \set{p_2},\top)$, in Fig. \ref{fig:refinement}
(b).

By firing the event $\mathbf{a}_1$ from this state, we reach
$(\tuple{q_1^1,q_1^2}, \emptyset,v=1)$, which is, again, accepting. We
check whether the path $\mathbf{init}.\mathbf{a}_2.\mathbf{a}_1$ is
feasible, which turns out not to be the case. For efficiency reasons,
we find the shortest suffix of this path that can be proved
infeasible. It turns out that the sequence $\mathbf{a}_2.\mathbf{a}_1$
is infeasible starting from the state $(\tuple{q_1^1,q_1^2},
\set{p_1},v=1)$, which is called the \emph{pivot}. This proof of
infeasibility yields the interpolant $\tuple{v=1,\Delta<x}$, and a new
predicate $\Delta<x$ is associated with $(\tuple{q_1^1,q_1^2},
\set{p_2})$. The refinement phase rebuilds only the subtree rooted
at the pivot state, in Fig.  \ref{fig:refinement} (b).

The procedure then builds the tree on Fig. \ref{fig:refinement} (c)
starting from the pivot node and finds the accepting state
$(\tuple{q_1^1,q_1^2}, \emptyset,\Delta<x)$ as the result of firing
the sequence $\mathbf{init}.\mathbf{a}_2.\mathbf{a}_2$. This path is
spurious, and the new predicate $v=2$ is associated with the location
$(\tuple{q_1^1,q_1^2}, \set{p_2})$. The pivot node is the same as in
Fig. \ref{fig:refinement} (b), and, by recomputing the subtree rooted
at this node with the new predicates, we obtain the tree in
Fig. \ref{fig:refinement} (d), in which all frontier nodes are
subsumed by their predecessors. Thus, no new event needs to be fired,
and the procedure can stop reporting that the trace inclusion holds.

\subsection{Related Work}

The trace inclusion problem has been previously addressed in the
context of timed automata \cite{ouaknine-worrell-lics04}. Although the
problem is undecidable in general, decidability is recovered when the
left-hand side automaton has at most one clock, or the only constant
appearing in the clock constraints is zero. These are essentially the
only known decidable cases of language inclusion for timed automata.

The study of \emph{data automata}
\cite{Bojanczyk:2011:TLD,habermehl-data} usually deals with the
complexity of decision problems in logics describing data languages
for simple theories, typically infinite data domains with equality.
Here we provide a semi-decision procedure for the language inclusion
between data automata controlled by generic first-order theories,
whose language-theoretic problems are undecidable.

Data words are also studied in the context of \emph{symbolic visibly pushdown
automata} (SVPA) \cite{symbVisPushDown:CAV14}. Language inclusion is decidable
for SVPAs with transition guards from a~decidable theory because SVPAs are
closed under complement and the emptiness can be reduced to a finite number of
queries expressible in the underlying theory of guards. Decidability comes here
at the cost of reducing the expressivity and forbidding comparisons between
adjacent positions in the input (only comparisons between matching call/return
positions of the input nested words are allowed).

Finally, several works on model checking infinite-state systems against CTL
\cite{existQuantHornCl:CAV13} and CTL* \cite{infStCTLstarVer:CAV15}
specifications are related to our problem as they check inclusion between the
set of computation trees of an infinite-state system and the set of trees
defined by a branching temporal logic specification. The verification of
existential CTL~formulae~\cite{existQuantHornCl:CAV13} is reduced to solving
forall-exists quantified Horn clauses by applying counterexample guided
refinement to discover witnesses for existentially quantified variables. The
work~\cite{infStCTLstarVer:CAV15} on CTL* verification of infinite systems is
based on partial symbolic determinization, using prophecy variables to summarize
the future program execution. For finite-state systems, automata are a strictly
more expressive formalism than temporal logics\footnote{For (in)finite words,
the class of LTL-definable languages coincides with the star-free languages,
which are a strict subclass of ($\omega$-)regular languages.}. Such a comparison
is, however, non-trivial for infinite-state systems. Nevertheless, we found the
data automata considered in this paper to be a natural tool for specifying
verification conditions of array programs \cite{lia,lpar08,cav09} and regular
properties of timed languages \cite{AlurDill94}.

\section{Preliminary Definitions}

Let $\nat$ denote the set of non-negative integers including zero. For
any $k,\ell \in \nat$, $k \leq \ell$, we write $[k,\ell]$ for the set
$\set{k,k+1,\ldots,\ell}$. We write $\false$ and $\true$ for the
boolean constants \emph{false} and \emph{true}, respectively. Given a
possibly infinite data domain $\mathcal{D}$, we denote by
$\thd=\tuple{\mathcal{D}, f_1,\ldots,f_m}$ the set of syntactically
correct first-order formulae with function symbols $f_1,\ldots,f_m$. A
variable $x$ is said to be \emph{free} in a~formula $\phi$, denoted as
$\phi(x)$, iff it does not occur under the scope of a quantifier.

Let $\vec{x} = \set{x_1,\ldots,x_n}$ be a finite set of variables. A
\emph{valuation} $\nu : \vec{x} \rightarrow \mathcal{D}$ is an
assignment of the variables in $\vec{x}$ with values from
$\mathcal{D}$. We denote by $\mathcal{D}^\vec{x}$ the set of such
valuations. For a formula $\phi(\vec{x})$, we denote by $\nu
\modelsthd \phi$ the fact that substituting each variable $x \in
\vec{x}$ by $\nu(x)$ yields a valid formula in the theory $\thd$. In
this case, $\nu$ is said to be a~\emph{model} of $\phi$. A formula is
said to be {\em satisfiable} iff it has a model. For a~formula
$\phi(\vec{x}, \vec{x'})$ where $\vec{x'} = \set{x' ~|~ x \in
  \vec{x}}$ and two valuations $\nu,\nu' \in \mathcal{D}^{\vec{x}}$,
we denote by $(\nu,\nu') \modelsthd \phi$ the fact that the formula
obtained from $\phi$ by substituting each $x$ with $\nu(x)$ and each
$x'$ with $\nu'(x')$ is valid in $\thd$.

\subsubsection{Data Automata.}

\emph{Data Automata} (DA) are extensions of non-deterministic finite
automata with variables ranging over an infinite data domain
$\mathcal{D}$, equipped with a first order theory $\thd$. Formally, a
DA is a tuple $A =\tuple{\mathcal{D}, \Sigma, \vec{x}, Q, \iota, F,
  \prod}$, where:
\begin{compactitem}
\item $\Sigma$ is a finite alphabet of input events and $\diamond
  \in \Sigma$ is a special padding symbol,
\item $\vec{x} = \set{x_1,\ldots,x_n}$ is a set of variables,
\item $Q$ is a finite set of {\em states}, $\iota\in Q$ is an
  \emph{initial} state, $F \subseteq Q$ are {\em final} states, and
\item $\prod$ is a set of {\em rules} of the form $q
  \arrow{\sigma,\phi(\vec{x}, \vec{x'})} q'$ where $\sigma\in\Sigma$
  is an alphabet symbol and $\phi(\vec{x},\vec{x'})$ is a formula in
  $\thd$.
\end{compactitem}
A \emph{configuration} of $A$ is a pair $(q,\nu) \in Q \times
\mathcal{D}^{\vec{x}}$. We say that a configuration $(q',\nu')$ is 
a~\emph{successor} of $(q,\nu)$ if and only if there exists a rule $q
\arrow{\sigma,\phi}{} q' \in \prod$ and $(\nu,\nu') \modelsthd
\phi$. We denote the successor relation by $(q,\nu)
\arrow{\sigma,\phi}{}_A (q',\nu')$, and we omit writing $\phi$ and $A$
when no confusion may arise. We denote by 
$\mathit{succ}_A(q,\nu) = \{(q',\nu') \mid (q,\nu) \arrow{}{}_A
(q',\nu')\}$ the set of successors of a configuration $(q,\nu)$.

A \emph{trace} is a finite sequence $w=(\nu_0,\sigma_0), \ldots,
(\nu_{n-1},\sigma_{n-1}),(\nu_n,\diamond)$ of pairs $(\nu_i,\sigma_i)$
taken from the infinite alphabet $\mathcal{D}^{\vec{x}} \times
\Sigma$. A \emph{run} of $A$ over the \emph{trace} $w$ is a sequence
of configurations $\pi : (q_0,\nu_0) \arrow{\sigma_0}{} (q_1,\nu_1)
\arrow{\sigma_1}{} \ldots \arrow{\sigma_{n-1}}{} (q_n,\nu_n)$. We say
that the run $\pi$ is \emph{accepting} if and only if $q_n \in F$, in
which case $A$ \emph{accepts} $w$. The \emph{language} of $A$, denoted
$\mathcal{L}(A)$, is the set of traces accepted by $A$.

\subsubsection{Data Automata Networks.}

A \emph{data automata network} (DAN) is a non-empty tuple $\mathcal{A}
= \tuple{A_1,\ldots,A_N}$ of data automata $A_i =
\tuple{\mathcal{D},\Sigma_i,\vec{x}_i,Q_i,\iota_i,F_i,\prod_i}$,
$i\in[1,N]$ whose sets of states are pairwise disjoint. A DAN is a
succint representation of an exponentially larger DA
$\mathcal{A}^e=\tuple{\mathcal{D},\Sigma_{\mathcal{A}},
  \vec{x}_{\mathcal{A}}, Q_{\mathcal{A}}, \iota_{\mathcal{A}},
  F_{\mathcal{A}}, \prod_{\mathcal{A}}}$, called the \emph{expansion}
of $\mathcal{A}$, where:
\begin{compactitem}
  \item $\Sigma_{\mathcal{A}} = \Sigma_1 \cup \ldots \cup \Sigma_N$ and
  $\vec{x}_{\mathcal{A}} = \vec{x}_1 \cup \ldots \cup \vec{x}_N$,
  \item $Q_{\mathcal{A}} = Q_1 \times \ldots \times Q_N$, $\iota_{\mathcal{A}} =
  \tuple{\iota_1, \ldots, \iota_N}$ and $F_{\mathcal{A}} = F_1 \times \ldots \times
  F_N$, 
  \item $\tuple{q_1, \ldots, q_N} \arrow{\sigma,\varphi}{} \tuple{q'_1, \ldots,
  q'_N}$ if and only if \begin{inparaenum}[(i)] \item for all $i \in I$, $q_i
  \arrow{\sigma,\varphi_i}{} q'_i$, \item for all $i \not\in I$, $q_i = q'_i$,
  and \item $\varphi \equiv \bigwedge_{i\in I} \varphi_i \wedge \bigwedge_{j
  \not\in I} \tau_j$, 
\end{inparaenum} where $I = \{i \in [1,N] \mid q_i \arrow{\sigma,\varphi_i}{}
q'_i \in \prod_i\}$ is the set of DA that can move from $q_i$ to
$q'_i$ while reading the input symbol $\sigma$, and $\tau_j \equiv
\bigwedge_{x \in \vec{x}_j \setminus \left(\bigcup_{i \in I}
  \vec{x}_i\right)} x'=x$ propagates the values of the local variables
in $A_j$ that are not updated by $\set{A_i}_{i\in I}$.
\end{compactitem} 
Intuitively, all automata that can read an input symbol synchronize
their actions on that symbol whereas the rest of the automata make an
idle step and copy the values of their local variables which are not
updated by the active automata. The language of the DAN $\mathcal{A}$
is defined as the language of its expansion DA,
i.e.\ $\mathcal{L}(\mathcal{A}) = \mathcal{L}(\mathcal{A}^e)$.

%
%

\subsubsection{Trace Inclusion.}

Let $\mathcal{A}$ be a DAN and
$\mathcal{A}^e=\tuple{\mathcal{D},\Sigma, \vec{x}_{\mathcal{A}},
  Q_{\mathcal{A}}, \iota_{\mathcal{A}}, F_{\mathcal{A}},
  \prod_{\mathcal{A}}}$ be its expansion. For a set of variables
$\vec{y}\subseteq\vec{x}_{\mathcal{A}}$, we denote by $\nu
\proj_{\vec{y}}$ the restriction of a~valuation $\nu \in
\mathcal{D}^{\vec{x}_{\mathcal{A}}}$ to the variables in
$\vec{y}$. For a trace \(w=(\nu_0,\sigma_0), \ldots,
(\nu_n,\diamond)\in\left(\mathcal{D}^{\vec{x}_{\mathcal{A}}} \times
\Sigma_{\mathcal{A}}\right)^*\), we denote by \(w \proj_{\vec{y}}\)
the trace \((\nu_0\proj_{\vec{y}},\sigma_0), \ldots,
(\nu_{n-1}\proj_{\vec{y}},\sigma_{n-1}),
(\nu_n\proj_{\vec{y}},\diamond) \in \left(\mathcal{D}^{\vec{y}} \times
\Sigma\right)^*\). We lift this notion to sets of words in the natural
way, by defining $\mathcal{L}(\mathcal{A}) \proj_{\vec{y}} = \set{w
  \proj_{\vec{y}} \mid w \in \mathcal{L}(\mathcal{A})}$.

We are now ready to define the trace inclusion problem on which we
focus in this paper. Given a DAN $\mathcal{A}$ as before and a DA
$B=\tuple{\mathcal{D},\Sigma,\vec{x}_B,Q_B,\iota_B,F_B,\prod_B}$ such
that $\vec{x}_B \subseteq \vec{x}_{\mathcal{A}}$, the \emph{trace
  inclusion problem} asks whether $\mathcal{L}(\mathcal{A})
\proj_{\vec{x}_B} \subseteq \mathcal{L}(B)$? The right-hand side DA
$B$ is called \emph{observer}, and the variables in $\vec{x}_B$ are
called \emph{observable} variables.

\section{Boolean Closure Properties of Data Automata} \label{sec:boolean-closure}

We show first that data automata are closed under the boolean
operations of union, intersection and complement and that they are
amenable to determinisation. Clearly, the emptiness problem is, in
general, undecidable, due to the result of Minsky on 2-counter
machines with integer variables, increment, decrement and zero test
\cite{minsky67}.

Let $A = \tuple{\mathcal{D}, \Sigma, \vec{x}, Q, \iota, F, \prod}$ be
a DA for the rest of this section. $A$ is said to be
\emph{deterministic} if and only if, for each trace $w \in
\mathcal{L}(A)$, $A$ has at most one run over $w$. The first result of
this section is that, interestingly, any DA can be determinized while
preserving its language. The determinisation procedure is a
generalization of the classical subset construction for Rabin-Scott
word automata on finite alphabets. The reason why determinisation is
possible for automata over an infinite data alphabet
$\mathcal{D}^{\vec{x}} \times \Sigma$ is that the successive values
taken by \emph{each variable} $x\in\vec{x}$ are tracked by the
language $\mathcal{L}(A) \subseteq \left(\mathcal{D}^{\vec{x}} \times
\Sigma\right)^*$. This assumption is crucial: a typical example of
automata over an infinite alphabet, that cannot be determinized, are
timed automata \cite{AlurDill94}, where only the elapsed time is
reflected in the language, and not the values of the variables
(clocks).

Formally, the \emph{deterministic} DA accepting the language
$\mathcal{L}(A)$ is defined as $A^d = \tuple{\mathcal{D}, \Sigma,
  \vec{x}, Q^d, \iota^d, F^d, \prod^d}$, where $Q^d=2^Q$,
$\iota^d=\set{\iota}$, $F^d=\set{P \subseteq Q \mid P \cap F \neq
  \emptyset}$ and $\prod^d$ is the set of rules $P
\arrow{\sigma,\theta}{} P'$ such that:
\begin{compactitem}
\item for all $p' \in P'$ there exists $p \in P$ and a rule 
  $p \arrow{\sigma,\psi}{} p' \in \prod$, 
\item \(\theta(\vec{x},\vec{x}') \equiv 
  \bigwedge_{\scriptscriptstyle{p' \in P'}}
  \bigvee_{\begin{array}{l}
      \scriptscriptstyle{p \arrow{\sigma,\psi}{} p' \in \prod} \\[-2mm]
      \scriptscriptstyle{p \in P}
  \end{array}}
  \psi \wedge 
  \bigwedge_{\scriptscriptstyle{p' \in Q \setminus P'}}
  \bigwedge_{\begin{array}{l}  
      \scriptscriptstyle{p \arrow{\sigma,\varphi}{} p' \in \prod} \\[-2mm]
      \scriptscriptstyle{p \in P}
      \end{array}}\neg\varphi\enspace.\)
\end{compactitem}
The main difference with the classical subset construction for
Rabin-Scott automata is that here we consider \emph{all sets} $P'$ of
states that have a predecessor in $P$, not just the maximal such set.
This refined subset construction takes into account not just the
alphabet symbols in $\Sigma$, but also the valuations of the variables
in $\vec{x}$. Observe, moreover, that $A^d$ can be built for any
theory $\thd$ that is closed under conjunction and negation. The
following lemma states the main properties of $A^d$.

\begin{lemma}\label{lemma:determinism}
  Given a DA $A = \tuple{\mathcal{D}, \Sigma, \vec{x}, Q, \iota, F, \prod}$, 
  \begin{inparaenum}[(1)]
  \item\label{it1:determinism} for any $w \in
    \left(\mathcal{D}^{\vec{x}} \times \Sigma\right)^*$ and $P \in
    Q^d$, $A^d$ has exactly one run on $w$ that starts in $P$, and
  \item\label{it2:determinism} $\mathcal{L}(A)=\mathcal{L}(A^d)$.
  \end{inparaenum}
\end{lemma}
\proof{ (\ref{it1:determinism}) Let $w = (\nu_0,\sigma_0), \ldots,
  (\nu_{n-1},\sigma_{n-1}), (\nu_n,\diamond)$ be an arbitrary trace
  and $P \subseteq Q$ be a state of $A^d$. We first build a run $\pi =
  (P_0,\nu_0) \arrow{\sigma_0,\theta_0}{} (P_1,\nu_1) \ldots
  \arrow{\sigma_{n-1},\theta_{n-1}}{} (P_n,\nu_n)$ of $A^d$ such that
  $P_0 = P$, by induction on $n\geq0$. If $n=0$, then
  $w=(\nu_0,\diamond)$ and $\pi=(P_0,\nu_0)$ is trivially a run of
  $A^d$ over $w$. For the induction step, let $n>0$ and suppose that
  $A^d$ has a run $(P_0,\nu_0) \arrow{\sigma_0,\theta_0}{} \ldots
  (P_{n-1},\nu_{n-1})$ such that $P_0=P$. We extend this run to a run
  over $w$, by considering:
  \[\begin{array}{rcl}
  P_n & = & \set{p \in Q \mid \exists q \in P_{n-1} ~.~ q
    \arrow{\sigma_{n-1},\phi}{} p \in \prod \text{ and } (\nu_{n-1},\nu_n)
    \modelsthd \phi} \\ 
  \theta_n & \equiv & \bigwedge_{\scriptscriptstyle{p' \in P_n}}
  \bigvee_{\begin{array}{l}
      \scriptscriptstyle{p \arrow{\sigma,\psi}{} p' \in \prod} \\[-2mm]
      \scriptscriptstyle{p \in P_{n-1}}
  \end{array}}
  \psi \wedge 
  \bigwedge_{\scriptscriptstyle{p' \in Q \setminus P_n}}
  \bigwedge_{\begin{array}{l}  
      \scriptscriptstyle{p \arrow{\sigma,\varphi}{} p' \in \prod} \\[-2mm]
      \scriptscriptstyle{p \in P_{n-1}}
      \end{array}}\neg\varphi \enspace. 
  \end{array}\]
  It is not hard to see that $(\nu_{n-1},\nu_n) \models \theta_n$,
  thus $(P_0,\nu_0) \arrow{\sigma_0,\theta_0}{} \ldots
  \arrow{\sigma_n,\theta_n} (P_n,\nu_n)$ is indeed a run of $A^d$ over
  $w$. To show that $\pi$ is unique, suppose, by contradiction, that
  there exists a different run $\pi' = (R_0,\nu_0)
  \arrow{\sigma_0,\omega_0}{} (R_1,\nu_1) \ldots
  \arrow{\sigma_{n-1},\omega_{n-1}}{} (R_n,\nu_n)$ such that
  $P_0=R_0=P$. Notice that the relation labeling any transition rule
  $P_i \arrow{\sigma_i,\theta_i}{} P_{i+1}$ is entirely determined by
  the sets $P_i$ and $P_{i+1}$, so two runs are different iff they
  differ in at least one state, i.e.\ and $P_j \neq R_j$, for some $j
  \in [1,n]$. Let $i$ denote the smallest such $j$ and suppose that
  there exists $p \in P_i$ such that $p \not\in R_i$ (the symmetrical
  case $p \in R_i$ and $p \not\in P_i$ is left to the reader). By the
  definition of $\prod^d$, there exists $q \in P_{i-1}=R_{i-1}$ such
  that $q \arrow{\sigma_{i-1},\psi}{} p \in \prod$. Since
  $(\nu_{i-1},\nu_i) \models \theta_{i-1} \wedge \omega_{i-1}$, we
  obtain that $(\nu_{i-1},\nu_i) \models \bigvee\{\psi \mid q
  \arrow{\sigma_{i-1},\psi}{} p \in \prod,~ q \in P_{i-1}\}$ and
  $(\nu_{i-1},\nu_i) \models \bigwedge\{\neg\psi \mid q
  \arrow{\sigma_{i-1},\psi}{} p \in \prod,~ q \in P_{i-1}\}$,
  contradiction. Thus $\pi$ is the only run of $A^d$ over $w$,
  starting in $P$.

  (\ref{it2:determinism}) Let $w = (\nu_0,\sigma_0), \ldots,
  (\nu_{n-1},\sigma_{n-1}), (\nu_n,\diamond)$ be a
  trace. ``$\subseteq$'' If $w \in \mathcal{L}(A)$, then $A$ has a run
  $(q_0,\nu_0) \arrow{\sigma_0,\phi_0}{} \ldots
  \arrow{\sigma_{n-1},\phi_{n-1}}{} (q_n,\nu_n)$ such that $q_0=\iota$
  and $q_n\in F$. By point (\ref{it1:determinism}), $A^d$ has a unique
  run $(P_0,\nu_0) \arrow{\sigma_0,\theta_0}{} \ldots
  \arrow{\sigma_{n-1},\theta_{n-1}}{} (P_n,\nu_n)$ over $w$. We prove
  that $q_i \in P_i$, by induction on $i\in[0,n]$. For $i=0$, we have
  $P_0=\set{\iota}$, by the definition of $A^d$. For the induction
  step, suppose that $i \in [1,n]$ and $q_{i-1} \in P_{i-1}$. By
  contradiction, assume that $q_i \not\in P_i$. Since
  $(\nu_{i-1},\nu_i) \modelsthd \theta_{i-1}$, we obtain
  $(\nu_{i-1},\nu_i) \modelsthd \neg\phi_{i-1}$, contradiction. Thus
  $q_i \in P_i$ for all $i \in [0,n]$, and $q_n \in P_n$, hence $P_n
  \cap F \neq \emptyset$. Then $P_n \in F^d$, and $w \in
  \mathcal{L}(A^d)$. ``$\supseteq$'' If $w \in \mathcal{L}(A^d)$, then
  $A^d$ has a (unique) run $(P_0,\nu_0) \arrow{\sigma_0,\theta_0}{}
  (P_1,\nu_1) \ldots \arrow{\sigma_{n-1},\theta_{n-1}}{} (P_n,\nu_n)$
  over $w$, such that $P_0=\set{\iota}$ and $P_n\cap F \neq
  \emptyset$. Then there exists $p_n\in P_n\cap F$ and, by the
  definition of $A^d$, there exists $p_{n-1} \in P_{n-1}$ such that
  $p_{n-1} \arrow{\sigma_{n-1},\psi_{n-1}}{} p_n \in \prod$ and
  $(\nu_{n-1},\nu_n) \modelsthd \psi_{n-1}$. Continuing this argument
  backwards, we can find a run $(q_0,\nu_0) \arrow{\sigma_0,\psi_0}{}
  \ldots \arrow{\sigma_{n-1},\psi_{n-1}}{} (q_n,\nu_n)$ of $A$ over
  $w$, such that $q_i \in P_i$, for all $i \in [0,n]$. Since
  $P_0=\set{\iota}$ and $q_n \in F$, we obtain that $w \in
  \mathcal{L}(A)$. \qed}

The construction of a deterministic DA recognizing the language of $A$
is key to defining a DA that recognizes the complement of $A$. Let
$\overline{A} =
\tuple{\mathcal{D},\Sigma,\vec{x},Q^d,\iota^d,Q^d\setminus
  F^d,\prod^d}$. In other words, $\overline{A}$ has the same structure
as $A^d$, and the set of final states consists of those subsets that
contain no final state, i.e.\ $\set{P \subseteq Q \mid P \cap F =
  \emptyset}$. Using Lemma \ref{lemma:determinism}, it is not
difficult to show that $\mathcal{L}(\overline{A}) =
\left(\mathcal{D}^{\vec{x}} \times \Sigma\right)^* \setminus
\mathcal{L}(A)$.

Next, we show closure of DA under intersection. Let $B =
\tuple{\mathcal{D},\Sigma,\vec{x},Q',\iota',F',\prod'}$ be a DA and
define $A \times B = \tuple{\mathcal{D},\Sigma,\vec{x},Q \times
  Q',(\iota,\iota'), F \times F',\prod^\times}$, where $(q,q')
\arrow{\sigma,\varphi}{} (p,p') \in \prod^\times$ if and only if $q
\arrow{\sigma,\phi}{} p \in \prod$, $q' \arrow{\sigma,\psi}{} p' \in
\prod'$ and $\varphi \equiv \phi \wedge \psi$. It is easy to show that
$\mathcal{L}(A \times B) = \mathcal{L}(A) \cap \mathcal{L}(B)$. DA are
also closed under union, since $\mathcal{L}(A) \cup \mathcal{L}(B) =
\mathcal{L}(\overline{\overline{A} \times \overline{B}})$.


Let us turn now to the trace inclusion problem. The following lemma
shows that this problem can be effectively reduced to an equivalent
language emptiness problem. However, note that this reduction does not
work when the trace inclusion problem is generalized by removing the
condition $\vec{x}_B \subseteq \vec{x}_A$. In other words, if the
observer uses local variables not shared with the network\footnote{For
  timed automata, this is the case since the only shared variable is
  the time, and the observer may have local clocks.}, i.e.\ $\vec{x}_B
\setminus \vec{x}_A \neq \emptyset$, the generalized trace inclusion
problem $\mathcal{L}(A)\proj_{\vec{x}_A \cap \vec{x}_B} \subseteq
\mathcal{L}(B)\proj_{\vec{x}_A \cap \vec{x}_B}$ has a negative answer
iff \emph{there exists a~trace} $w = (\nu_0,\sigma_0), \ldots,
(\nu_n,\diamond) \in \mathcal{L}(A)$ such that, \emph{for all
  valuations} $\mu_0,\ldots,\mu_n\in
\mathcal{D}^{\vec{x}_B\setminus\vec{x}_A}$, we have
$w'=(\nu_0\proj_{\vec{x}_A \cap \vec{x}_B}\cup~\mu_0,\sigma_0),
\ldots, (\nu_n\proj_{\vec{x}_A \cap \vec{x}_B}\cup~\mu_n,\diamond)
\not\in \mathcal{L}(B)$. This kind of quantifier alternation cannot be
easily accommodated within the framework of language emptiness, in
which only one type of (existential) quantifier occurs.

\begin{lemma}\label{lemma:inclusion-emptiness}
  Given DA $A =
  \tuple{\mathcal{D},\Sigma,\vec{x}_A,Q_A,\iota_A,F_A,\prod_A}$ and $B
  = \tuple{\mathcal{D},\Sigma,\vec{x}_B,Q_B,\iota_B,F_B,\prod_B}$ such
  that $\vec{x}_B \subseteq \vec{x}_A$. Then
  $\mathcal{L}(A)\proj_{\vec{x}_B} \subseteq \mathcal{L}(B)$ if and only if 
  $\mathcal{L}(A \times \overline{B}) = \emptyset$. 
\end{lemma}
\proof{ We have $\mathcal{L}(A)\proj_{\vec{x}_B} \subseteq
  \mathcal{L}(B)$ iff $\mathcal{L}(A)\proj_{\vec{x}_B} \cap
  \mathcal{L}(\overline{B}) = \mathcal{L}(A \times
  \overline{B})\proj_{\vec{x}_B} = \emptyset$ iff $\mathcal{L}(A
  \times \overline{B}) = \emptyset$. \qed}

The trace inclusion problem is undecidable, which can be shown by
reduction from the language emptiness problem for DA (take $B$ such
that $\mathcal{L}(B)=\emptyset$). However the above lemma shows that
any semi-decision procedure for the language emptiness problem can
also be used to deal with the trace inclusion problem.

\section{Abstract, Check, and Refine for Trace Inclusion}
\label{sec:algorithm}

This section describes our semi-algorithm for checking the trace
inclusion between a~given network $\mathcal{A}$ and an observer
$B$. Let $\mathcal{A}^e$ denote the expansion of $\mathcal{A}$,
defined in the previous. In the light of Lemma
\ref{lemma:inclusion-emptiness}, the trace inclusion problem
$\mathcal{L}(\mathcal{A})\proj_{\vec{x}_B} \subseteq \mathcal{L}(B)$,
where the set of observable variables $\vec{x}_B$ is included in the
set of network variables, can be reduced to the language emptiness
problem $\mathcal{L}(\mathcal{A}^e \times \overline{B}) = \emptyset$.

Although language emptiness is undecidable for data automata
\cite{minsky67}, several cost-effective semi-algorithms and tools
\cite{lazy-abstraction,mcmillan06,rybal-pldi11,fast} have been
developped, showing that it is possible, in many practical cases, to
provide a yes/no answer to this problem. However, to apply one of the
existing off-the-shelf tools to our problem, one needs to build the
product automaton $\mathcal{A}^e \times \overline{B}$ prior to the
analysis. Due to the inherent state explosion caused by the
interleaving semantics of the network as well as by the
complementation of the observer, such a solution would not be
efficient in practice.

To avoid building the product automaton, our procedure builds
\emph{on-the-fly} an over-approximation of the (possibly infinite) set
of reachable configurations of $\mathcal{A}^e \times
\overline{B}$. This over-approximation is defined using the approach
of \emph{lazy predicate abstraction} \cite{lazy-abstraction}, combined
with \emph{counterexample-driven abstraction refinement} using
\emph{interpolants} \cite{mcmillan06}. We store the explored abstract
states in a structure called an \emph{antichain tree}. In general,
antichain-based algorithms
\cite{henzinger06} store only states
which are incomparable w.r.t. a partial order called
\emph{subsumption}. Our method can be thus seen as an extension of the
antichain-based language inclusion algorithm \cite{abdulla} to infinite
state systems by means of predicate abstraction and
interpolation-based refinement. Since the trace inclusion problem is
undecidable in general, termination of our procedure is not
guaranteed; in the following, we shall, however, call our procedure an
algorithm for the sake of brevity.

\vspace*{-1mm}\subsection{Antichain Trees}

In this section, we define antichain trees, which are the main data
structure of the trace inclusion (semi-)algorithm. Let $\mathcal{A} =
\tuple{A_1,\ldots,A_N}$ be a network of automata where $A_i =
\tuple{\mathcal{D},\Sigma_i,\vec{x}_i,Q_i,\iota_i,F_i,\prod_i}$, for
all $i \in [1,N]$, and let $B =
\tuple{\mathcal{D},\Sigma,\vec{x}_B,Q_B,\iota_B,F_B,\prod_B}$ be an
observer such that $\vec{x}_B \subseteq \bigcup_{i=1}^N\vec{x}_i$. We
also denote by $\mathcal{A}^e = \tuple{\mathcal{D},
  \Sigma_{\mathcal{A}}, \vec{x}_{\mathcal{A}}, Q_{\mathcal{A}},
  \iota_{\mathcal{A}}, F_{\mathcal{A}}, \prod_{\mathcal{A}}}$ the
expansion of the network $\mathcal{A}$ and by $\mathcal{A}^e \times
\overline{B} = \tuple{\mathcal{D}, \Sigma_{\mathcal{A}},
  \vec{x}_{\mathcal{A}}, Q^p, \iota^p, F^p, \prod^p}$ the product
automaton used for checking language inclusion.

An \emph{antichain tree} for the network $\mathcal{A}$ and the
observer $B$ is a tree whose nodes are labeled by \emph{product
  states} (see Fig. \ref{fig:refinement} for examples). Intuitively, a
product state is an over-approximation of the set of configurations of
the product automaton $\mathcal{A}^e \times \overline{B}$ that share
the same control state. Formally, a \emph{product state for
  $\mathcal{A}$ and $B$} is a tuple $s=(\vec{q},P,\Phi)$
where\begin{inparaenum}[(i)]
\item $(\vec{q},P)$ is a~state of $\mathcal{A}^e \times \overline{B}$
  with $\vec{q}=\tuple{q_1,\ldots,q_N}$ being a state of the network
  expansion $\mathcal{A}^e$ and $P$ being a set of states of the
  observer $B$, and
\item $\Phi(\vec{x}_{\mathcal{A}}) \in \thd$ is a formula which
  defines an over-approximation of the set of valuations of the
  variables $\vec{x}_{\mathcal{A}}$ that reach the state
  $(\vec{q},P)$ in $\mathcal{A}^e \times \overline{B}$.
\end{inparaenum}
A product state $s=(\vec{q},P,\Phi)$ is a finite representation of a
possibly infinite set of configurations of 
$\mathcal{A}^e \times \overline{B}$, denoted as $\sem{s} =
\{(\vec{q},P,\nu) \mid \nu \modelsthd \Phi\}$.


To build an over-approximation of the set of reachable states of the product
automaton, we need to compute, for a product state $s$, an over-approximation of
the set of configurations that can be reached in one step from $s$. To this end,
we define first a finite abstract domain of product states, based on the notion
of \emph{predicate map}. A predicate map is a partial function that associates
sets of facts about the values of the variables used in the product automaton,
called \emph{predicates}, with components of a product state, called
\emph{substates}. The reason behind the distribution of predicates over
substates is two-fold. First, we would like the abstraction to be \emph{local},
i.e.\ the predicates needed to define a certain subtree in the antichain must be
associated with the labels of that subtree only. Second, once a predicate
appears in the context of a substate, it should be subsequently reused whenever
that same substate occurs as part of another product state.

Formally, a \emph{substate} of a state $(\tuple{q_1,\ldots,q_N},P) \in
Q^p$ of the product automaton $\mathcal{A}^e \times \overline{B}$ is a
pair $(\tuple{q_{i_1},\ldots,q_{i_k}},S)$ such
that \begin{inparaenum}[(i)]
\item $\tuple{q_{i_1},\ldots,q_{i_k}}$ is a subsequence of
  $\tuple{q_1,\ldots,q_N}$, and
\item $S \neq \emptyset$ only if $S \cap P \neq \emptyset$. 
\end{inparaenum}
We denote the substate relation by $(\tuple{q_{i_1},\ldots,q_{i_k}},S)
\vartriangleleft (\tuple{q_1,\ldots,q_N},P)$. The substate relation
requires the automata $A_{i_1}, \ldots, A_{i_k}$ of the network
$\mathcal{A}$ to be in the control states $q_{i_1},\ldots,q_{i_k}$
simultaneously, and the observer $B$ to be in at least some state of
$S$ provided $S\neq\emptyset$ (if $S=\emptyset$, the state of $B$ is
considered to be irrelevant). Let $\mathcal{S}_{\tuple{\mathcal{A},B}}
= \set{r \mid \exists q \in Q^p ~.~ r \vartriangleleft q}$ be the set
of substates of a state of $\mathcal{A}^e \times \overline{B}$.

A \emph{predicate map} $\Pi : \mathcal{S}_{\tuple{\mathcal{A},B}}
\rightarrow 2^{\smallthd}$ associates each substate $(\vec{r},S) \in
Q_{i_1} \times \ldots \times Q_{i_k} \times 2^{Q_B}$ with a set of
formulae $\pi(\vec{x})$ where \begin{inparaenum}[(i)]
\item $\vec{x} = \vec{x}_{i_1} \cup \ldots \cup \vec{x}_{i_k} \cup
  \vec{x}_B$ if $S \neq \emptyset$, and
\item $\vec{x} = \vec{x}_{i_1} \cup \ldots \cup \vec{x}_{i_k}$ if $S =
  \emptyset$.
\end{inparaenum}
Notice that a predicate associated with a substate refers only to the
local variables of those network components $A_{i_1}, \ldots,
A_{i_k}$ and of the observer $B$ that occur in the particular
substate.

\begin{example}\label{ex:predicate-map} 
The antichain in Fig. \ref{fig:refinement} (d) uses the predicate map
\((\langle q_1^1,q_1^2\rangle,\{p_1\}) \mapsto \set{v=1}\),
\((\langle q_1^1,q_1^2\rangle,\{p_2\}) \mapsto \set{\Delta<x,
  v=2}\). $\blacksquare$
\end{example}

We are now ready to define the abstract semantics of the product
automaton $\mathcal{A}^e \times \overline{B}$, induced by a given
predicate map. For convenience, we define first a set
$\mathit{Post}(s)$ of \emph{concrete successors} of a product state
$s=(\vec{q},P,\Phi)$ such that $(\vec{r},S,\Psi) \in \mathit{Post}(s)$ if and
only if\begin{inparaenum}[(i)]
\item the product automaton $\mathcal{A}^e \times \overline{B}$ has a
  rule $(\vec{q},P) \arrow{\sigma,\theta}{} (\vec{r},S) \in \Delta^p$ and
$\Psi(\vec{x}_{\mathcal{A}}) \equiv \exists
  \vec{x}'_{\mathcal{A}} ~.~ \Phi(\vec{x}'_{\mathcal{A}}) \wedge
  \theta(\vec{x}'_{\mathcal{A}}, \vec{x}_{\mathcal{A}}) \not\rightarrow \false$.
\end{inparaenum}
The set of concrete successors does not contain states with empty set
of valuations because these states are unreachable in $\mathcal{A}^e
\times \overline{B}$.

Given a predicate map $\Pi$, the set $\mathit{Post}_\Pi(s)$ of
\emph{abstract successors} of a product state $s$ is defined as
follows: $(\vec{r},S,\Psi^\sharp) \in \mathit{Post}_\Pi(s)$ if and
only if \begin{inparaenum}[(i)]
\item there exists a product state $(\vec{r},S,\Psi) \in
  \mathit{Post}(s)$ and
\item $\Psi^\sharp(\vec{x}_{\mathcal{A}}) \equiv \bigwedge_{r
  \vartriangleleft (\vec{r},S)} \bigwedge \set{\pi \in \Pi(r) \mid
  \Psi \rightarrow \pi}$. 
\end{inparaenum}
In other words, the set of data valuations that are reachable by an
abstract successor is the tightest over-approximation of the concrete
set of reachable valuations, obtained as the conjunction of the
available predicates from the predicate map that over-approximate
this set.

\begin{example}(\emph{contd. from Ex. \ref{ex:predicate-map}}) 
Consider the antichain from Fig. \ref{fig:refinement} (d). The
concrete successors of $s=(\tuple{q_1^1,q_1^2},\set{p_1},v=1)$ are
$(\tuple{q_1^1,q_1^2},\set{p_1},\Psi_1)$ and
$(\tuple{q_1^1,q_1^2},\set{p_2},\Psi_2)$:
\[\begin{array}{rcl}
\Psi_1 & \equiv & \exists v',x',\Delta' ~.~ v'=1 \wedge x=x'+1 \wedge v=1 
\wedge \Delta=\Delta' \wedge 0\leq x'< \Delta \wedge v=v' \\
\Psi_2 & \equiv & \exists v',x',\Delta' ~.~ v'=1 \wedge x=x'+1 \wedge v=2
\wedge \Delta=\Delta' \wedge \Delta\leq x'< 2\Delta \wedge v=v'+1\vspace*{-1mm}
\end{array}\]
\vspace*{-1mm}With predicate map $\Pi$ from Ex. \ref{ex:predicate-map}, 
$\mathit{Post}_\Pi(s) =
\{(\tuple{q_1^1,q_1^2},\set{p_1},\Psi^\sharp_1),(\tuple{q_1^1,q_1^2},\set{p_2},\Psi^\sharp_2)\}$:
\[\begin{array}{lcl}
\Psi_1 \rightarrow v=1 & \Rightarrow & \Psi_1^\sharp \equiv v=1 \\
\Psi_2 \rightarrow v=2 \text{ and } \Psi_2 \rightarrow \Delta<x & 
\Rightarrow & \Psi_2^\sharp \equiv v=2 \wedge \Delta<x \enspace.\enspace\blacksquare
\end{array}\]
\end{example}

Finally, an \emph{antichain tree} (or, simply antichain) $\mathcal{T}$
for $\mathcal{A}$ and $B$ is a tree whose nodes are labeled with
product states and whose edges are labeled by input symbols and
concrete transition relations. Let $\nat^*$ be the set of finite
sequences of natural numbers that denote the positions in the
tree. For a tree position $p \in \nat^*$ and $i \in \nat$, the
position $p.i$ is a \emph{child} of $p$. A set $S \subseteq \nat^*$ is
said to be \emph{prefix-closed} if and only if, for each $p \in S$ and
each prefix $q$ of $p$, we have $q \in S$ as well. The root of the
tree is denoted by the empty sequence $\varepsilon$.

Formally, an antichain $\mathcal{T}$ is a set of pairs $\tuple{s,p}$,
where $s$ is a product state and $p\in\nat^*$ is a tree position, such
that\begin{inparaenum}[(1)]
\item for each position $p \in \nat^*$ there exists at most one
  product state $s$ such that $\tuple{s,p} \in \mathcal{T}$,
\item the set $\set{p \mid \tuple{s,p} \in \mathcal{T}}$ is
  prefix-closed, 
\item $(\mathit{root}_{\tuple{\mathcal{A},B}},\varepsilon) \in
  \mathcal{T}$ where
  $\mathit{root}_{\tuple{\mathcal{A},B}}=(\tuple{\iota_1,\ldots,\iota_N},
  \set{\iota_B}, \true)$ is the label of the root, and 
\item for each edge $(\tuple{s,p},\tuple{t,p.i})$ in $\mathcal{T}$,
  there exists a predicate map $\Pi$ such that $t \in
  \mathit{Post}_\Pi(s)$.
\end{inparaenum}
For the latter condition, if $s=(\vec{q},P,\Phi)$ and
$t=(\vec{r},S,\Psi)$, there exists a~unique rule $(\vec{q},P)
\arrow{\sigma,\theta}{} (\vec{r},S) \in \prod^p$, and we shall
sometimes denote the edge as $s \arrow{\sigma,\theta}{} t$ or simply
$s \arrow{\theta}{} t$ when the tree positions are not important.

Each antichain node $n=(s,d_1\ldots d_k) \in \mathcal{T}$ is naturally
associated with a path from the root to itself $\rho \colon n_0
\arrow{\sigma_1,\theta_1}{} n_1 \arrow{\sigma_2,\theta_2}{} \ldots
\arrow{\sigma_2,\theta_k}{} n_k$. We denote by $\rho_i$ the node $n_i$
for each $i\in[0,k]$, and by $\len{\rho}=k$ the length of the path.
The \emph{path formula} associated with $\rho$ is $\Theta(\rho) \equiv
\bigwedge_{i=1}^k
\theta(\vec{x}_{\mathcal{A}}^{i-1},\vec{x}_{\mathcal{A}}^i)$ where
$\vec{x}_{\mathcal{A}}^i = \set{x^i \mid x \in \vec{x}_{\mathcal{A}}}$
is a~set of indexed variables for each $i\in[0,k]$.

\begin{example} \label{ex:path-formula} Consider the path $\rho ~:~ (\tuple{q_0^1,q_0^2},\set{p_0},\top) 
  \arrow{\mathbf{init}}{} (\tuple{q_1^1,q_1^2},\set{p_1},v=1)
  \arrow{\mathbf{a_2}}{} (\tuple{q_1^1,q_1^2},\set{p_2}, \Delta<x)
  \arrow{\mathbf{a_2}}{} (\tuple{q_1^1,q_1^2},\emptyset,\Delta<x)$ in
  the antichain from Fig. \ref{fig:refinement} (c). The path formula
  of $\rho$ is $\Theta(\rho) \equiv \theta_1 \wedge \theta_2 \wedge
  \theta_3$ where:\vspace*{-1mm}
  \[\begin{array}{rcl}
  \theta_1 & \equiv & v_1=1 \wedge x_1=0 \wedge 0 <\Delta_1 \\
  \theta_2 & \equiv & v_2 = v_1+ 1 \wedge  \Delta_2 =\Delta_1 \wedge  
  v_2=2 \wedge x_2 = x_1 + 1 \wedge \Delta_1 \leq x_1 < 2 \Delta_1 \wedge \neg( v_2= v_1) \\
  \theta_3 & \equiv & v_3 = 2 \wedge  \Delta_3 =\Delta_2  \wedge x_3 = x_2+1 \wedge
  \Delta_2 \leq x_2 < 2 \Delta_2 \wedge \neg( v_3= v_2) \enspace. \enspace \blacksquare
  \end{array}\]
\end{example}

\subsection{Counterexample-driven Abstraction Refinement}\label{sec:cex}

A \emph{counterexample} is a path from the root of the antichain to a
node which is labeled by an \emph{accepting} product state. A product
state $(\vec{q},P,\Phi)$ is said to be \emph{accepting} iff
$(\vec{q},P)$ is an accepting state of the product automaton
$\mathcal{A}^e \times \overline{B}$, i.e.\ $\vec{q} \in
F_{\mathcal{A}}$ and $P \cap F_B = \emptyset$. A counterexample is
said to be \emph{spurious} if its path formula is unsatisfiable,
i.e.\ the path does not correspond to a concrete execution of
$\mathcal{A}^e \times \overline{B}$. In this case, we need
to \begin{inparaenum}[(i)]
\item remove the path $\rho$ from the current antichain and
\item refine the abstract domain in order to exclude the occurrence of
  $\rho$ from future state space exploration. 
\end{inparaenum}

Let $\rho : \mathit{root}_{\tuple{\mathcal{A},B}} =
(\vec{q}_0,P_0,\Phi_0) \arrow{\theta_1}{} (\vec{q}_1,P_1,\Phi_1)
\arrow{\theta_2}{} \ldots \arrow{\theta_k}{} (\vec{q}_k,P_k,\Phi_k)$
be a spurious counterexample in the following. For efficiency
reasons, we would like to save as much work as possible and remove
only the smallest suffix of $\rho$ which caused the spuriousness. For
some $j \in [0,k]$, let $\Theta^j(\rho) \equiv
\Phi_j(\vec{x}^0_{\mathcal{A}}) \wedge \bigwedge_{i=j}^k
\theta_i(\vec{x}^{i-j}_{\mathcal{A}}, \vec{x}^{i-j+1}_{\mathcal{A}})$
be the formula defining all sequences of data valuations that start in
the set $\Phi_j$ and proceed along the suffix $(\vec{q}_j,P_j,\Phi_j)
\arrow{}{} \ldots \arrow{}{} (\vec{q}_k,P_k,\Phi_k)$ of $\rho$. The
\emph{pivot} of a path $\rho$ is the maximal position $j\in[0,k]$ such
that $\Theta^j(\rho) = \false$, and $-1$ if $\rho$ is not spurious.


\begin{example}\label{ex:spurious-ce}(\emph{contd. from Ex. \ref{ex:path-formula}}) 
The path formula $\Theta(\rho) \equiv \theta_1 \wedge \theta_2 \wedge
\theta_3$ from Ex.  \ref{ex:path-formula} is unsatisfiable, thus
$\rho$ is a spurious counterexample. Moreover, we have $\Theta^1(\rho)
\equiv \true \wedge \theta_2 \wedge \theta_3 \rightarrow \false$
because of the unsatisfiable subformula $v_2=2 \wedge v_3 = 2 \wedge
\neg( v_3= v_2)$. Since $\Theta^2(\rho)$ is satisfiable, the pivot of
$\rho$ is $1$. $\blacksquare$ \end{example}

Finally, we describe the refinement of the predicate map, which
ensures that a given spurious counterexample will never be found in a
future iteration of the abstract state space exploration. The
refinement is based on the notion of \emph{interpolant}
\cite{mcmillan06}. 

\begin{definition}\label{def:interpolant}
  Given a formula $\Phi(\vec{x})$ and a sequence
  $\tuple{\theta_1(\vec{x},\vec{x}'), \ldots,
    \theta_k(\vec{x},\vec{x}')}$ of formulae, an \emph{interpolant} is
  a sequence of formulae $\mathbf{I} = \tuple{I_0(\vec{x}), \ldots,
    I_{k}(\vec{x})}$ where:%
  \begin{inparaenum}[(1)]
  \item $\Phi \rightarrow I_0$, 
  \item $I_k \rightarrow \false$, and
  \item $I_{i-1}(\vec{x}) \wedge \theta_i(\vec{x},\vec{x}')
    \rightarrow I_i(\vec{x}')$ for all $i \in [1,k]$.
  \end{inparaenum}
\end{definition}
\vspace*{-0.5mm}Any given interpolant is a witness for the unsatisfiability of a
(suffix) path formula $\Theta^j(\rho)$. Dually, if \emph{Craig's
  Interpolation Lemma} \cite{craig} holds for the considered
first-order data theory $\thd$, any infeasible path formula is
guaranteed to have an interpolant.

\begin{example}(\emph{contd. from Ex. \ref{ex:spurious-ce}}) 
The path formula $\Theta^1(\rho)$ in Ex. \ref{ex:spurious-ce} has
the interpolant $I=\tuple{\top,v=2,\bot }$. $\blacksquare$ \end{example}

Given a spurious counterexample $\rho$ with pivot $j\geq0$, an
interpolant $\mathbf{I}=\tuple{I_0, \ldots, I_{k-j}}$ for the
infeasible path formula $\Theta^j(\rho)$ can be used to refine the
abstract domain by augmenting the predicate map $\Pi$. As an effect of
this refinement, the antichain construction algorithm will avoid every
path with the suffix $(\vec{q}_j,P_j,\Phi_j) \arrow{}{} \ldots
\arrow{}{} (\vec{q}_k,P_k,\Phi_k)$ in a~future iteration. If $I_i \iff
C_i^1(\vec{y}_1) \wedge \ldots \wedge C_i^{m_i}(\vec{y}_{m_i})$ is a
conjunctive normal form (CNF) of the $i$-th component of the
interpolant, we consider the substate $(\vec{r}_i^\ell,S_i^\ell)$ for
each $C_i^\ell(\vec{y}_\ell)$ where $l \in [1,m_i]$:\begin{compactitem}
\item $\vec{r}_i^\ell = \tuple{q_{i_1}, \ldots, q_{i_h}}$ where $1
  \leq i_1 < \ldots < i_h \leq N$ is the largest sequence of indices
  such that $\vec{x}_{i_g} \cap \vec{y}_\ell \neq \emptyset$ for
  each $g\in[1,h]$ and the
  set $\vec{x}_{i_g}$ of variables of the network component DA $A_{i_g}$,
\item $S_i^\ell = P_j$ if $\vec{x}_B \cap \vec{y}_\ell \neq
  \emptyset$, and $S_i^\ell=\emptyset$, otherwise.
\end{compactitem} 

A predicate map $\Pi$ is said to be \emph{compatible} with a spurious
path $\rho : s_0 \arrow{\theta_1}{} \ldots \arrow{\theta_k}{} s_k$
with pivot $j\geq0$ if $s_j=(\vec{q}_j,P_j,\Phi_j)$ and there is an
interpolant $\mathbf{I}=\tuple{I_0,\ldots,I_{k-j}}$ of the suffix
$\tuple{\theta_1,\ldots,\theta_k}$ wrt. $\Phi_j$ such that, for each
clause $C$ of some equivalent CNF of $I_i$, $i \in [0,k-j]$, it holds
that $C\in \Pi(r)$ for some substate $r \vartriangleleft s_{i+j}$. The
following lemma proves that, under a predicate map compatible with a
spurious path $\rho$, the antichain construction will exclude further
paths that share the suffix of $\rho$ starting with its pivot.

\begin{lemma}\label{lemma:refinement}
  Let $\rho : (\vec{q}_0,P_0,\Phi_0) \arrow{\theta_0}{}
  (\vec{q}_1,P_1,\Phi_1) \arrow{\theta_1}{} \ldots
  \arrow{\theta_{k-1}}{} (\vec{q}_k,P_k,\Phi_k)$ be a spurious
  counterexample and $\Pi$ be a predicate map compatible with
  $\rho$. Then, there is no sequence of product states
  $(\vec{q}_j,P_j,\Psi_0), \ldots, (\vec{q}_k,P_k,\Psi_{k-j})$ such
  that:\begin{inparaenum}[(1)]
  \item $\Psi_0 \rightarrow \Phi_j$ and
  \item $(\vec{q}_{i+1},P_{i+1},\Psi_{i-j+1}) \in
  \mathit{Post}_\Pi((\vec{q}_i,P_i,\Psi_{i-j}))$ for all
  $i\in[j,k-1]$.
  \end{inparaenum}
\end{lemma}
\proof{ Let $j\in[0,k]$ be the pivot of $\rho$. Since $\rho$ is
  spurious, there exists an interpolant
  $\mathbf{I}=\tuple{I_0,\ldots,I_{k-j}}$ for $\Phi_j$ and
  $\tuple{\theta_j,\ldots,\theta_k}$. It is sufficient to prove that
  $\Psi_i\rightarrow I_i$ for all $i\in[0,k-j]$. Since $I_{k-j} =
  \false$, we obtain $\Psi_{k-j} = \false$, and consequently
  $(\vec{q}_{k-j},P_{k-j},\false) \in
  \mathit{Post}_\Pi((\vec{q}_{k-j-1},P_{k-j-1},\Psi_{k-j-1}))$. By the
  definition of $\mathit{Post}_\Pi$, we have
  $(\vec{q}_{k-j},P_{k-j},\false) \in
  \mathit{Post}((\vec{q}_{k-j-1},P_{k-j-1},\Psi_{k-j-1}))$, which
  contradicts with the definition of $\mathit{Post}$. We show that
  $\Psi_i\rightarrow I_i$ for all $i\in[0,k-j]$, by induction on
  $k-j$. For the base case $k-j=0$, we have $\Psi_0 \rightarrow \Phi_j
  \rightarrow I_0$. For the induction step, we assume
  $\Psi_i\rightarrow I_i$ for all $i\in[0,k-j-1]$ and prove
  $\Psi_{k-j}\rightarrow I_{k-j}$. By the induction hypothesis, we
  have: \[\begin{array}{rcl}
  \Psi_{k-j-1}(\vec{x}_{\mathcal{A}}) & \rightarrow & I_{k-j-1}(\vec{x}_{\mathcal{A}}) \\
  \Psi_{k-j-1}(\vec{x}_{\mathcal{A}}) \wedge \theta_{k-j-1}(\vec{x}_{\mathcal{A}},\vec{x}'_{\mathcal{A}}) & 
  \rightarrow & I_{k-j-1}(\vec{x}_{\mathcal{A}}) \wedge \theta_{k-j-1}(\vec{x}_{\mathcal{A}},\vec{x}'_{\mathcal{A}}) 
  \rightarrow I_{k-j}(\vec{x}'_{\mathcal{A}}) \enspace. 
  \end{array}\]
  Let $C_1 \wedge \ldots \wedge C_\ell$ be the CNF of $I_{k-j}$. Since
  $\Pi$ is compatible with $\rho$, for each clause $C_i$, there exists
  a substate $r \vartriangleleft (\vec{q}_k,P_k)$ such that $C_i \in
  \Pi(r)$. By the definition of $\mathit{Post}_\Pi$, we obtain that
  $\Psi_{k-j} \rightarrow C_i$ for each $i\in[1,\ell]$, hence
  $\Psi_{k-j} \rightarrow I_{k-j}$. 
  \qed}

Observe that the refinement induced by interpolation is \emph{local}
since $\Pi$ associates sets of predicates with substates of the states
in $\mathcal{A}^e \times \overline{B}$, and the update impacts only
the states occurring within the suffix of that particular spurious
counterexample.

\subsection{Subsumption}

The main optimization of antichain-based algorithms \cite{abdulla}
for checking language inclusion of automata over finite alphabets is
that product states that are \emph{subsets} of already visited states
are never stored in the antichain. On the other hand, language
emptiness semi-algorithms, based on \emph{predicate abstraction}
\cite{mcmillan06} use a similar notion to cover newly generated
abstract successor states by those that were visited sooner and that
represent larger sets of configurations. In this case, state coverage
does not only increase efficiency but also ensures termination of the
semi-algorithm in many practical cases.

In this section, we generalize the subset relation used in classical
antichain algorithms with the notion of coverage from predicate
abstraction, and we define a more general notion of \emph{subsumption}
for data automata. Given a state $(\vec{q},P)$ of the product
automaton $\mathcal{A}^e \times \overline{B}$ and a valuation $\nu \in
\mathcal{D}^{\vec{x}_{\mathcal{A}}}$, the \emph{residual language}
$\mathcal{L}_{(\vec{q},P,\nu)}(\mathcal{A}^e \times \overline{B})$ is
the set of traces $w$ accepted by $\mathcal{A}^e \times \overline{B}$
from the state $(\vec{q},P)$ such that $\nu$ is the first valuation
which occurs on $w$. This notion is then lifted to product states as
follows: $\mathcal{L}_s(\mathcal{A}^e \times \overline{B}) =
\bigcup_{(\vec{q},P,\nu)\in\sem{s}}
\mathcal{L}_{(\vec{q},P,\nu)}(\mathcal{A}^e \times \overline{B})$
where $\sem{s}$ is the set of configurations of the product automaton
$\mathcal{A}^e \times \overline{B}$ represented by the given product
state $s$.

\begin{definition}\label{def:subsumption}
  Given a DAN $\mathcal{A}$ and a DA $B$, a partial order
  $\sqsubseteq$ is a \emph{subsumption} provided that, for any two
  product states $s$ and $t$, we have $s \sqsubseteq t$ only if
  $\mathcal{L}_s(\mathcal{A}^e \times \overline{B}) \subseteq
  \mathcal{L}_t(\mathcal{A}^e \times \overline{B})$.
\end{definition}

A procedure for checking the emptiness of $\mathcal{A}^e \times \overline{B}$
needs not continue the search from a product state $s$ if it has already visited
a product state $t$ that subsumes $s$. The intuition is that any counterexample
discovered from $s$ can also be discovered from $t$. The trace inclusion
semi-algorithm described below in Section~\ref{sec:trace-inclusion-algorithm}
works, in principle, with any given subsumption relation. In practice, our
implementation uses the subsumption relation defined by the lemma below:

\begin{lemma}\label{lemma:img-subsumption} The relation defined s.t.
$(\vec{q},P,\Phi) \sqsubseteq_{\mathit{img}} (\vec{r},S,\Psi) \iff
\vec{q}=\vec{r},~ P \supseteq S \text{, and } \Phi \rightarrow \Psi$ is a
subsumption.\end{lemma}
\proof{ For any valuation $\nu \in
  \mathcal{D}^{\vec{x}_{\mathcal{A}}}$, we have
  $\mathcal{L}_{(\vec{q},P,\nu)}(\mathcal{A}^e \times \overline{B}) =
  \mathcal{L}_{(\vec{q},\nu)}(\mathcal{A}^e) \cap
  \mathcal{L}_{(P,\nu)}(\overline{B})$. Since $P \supseteq S$, we have
  $\mathcal{L}_{(P,\nu)}(B) \supseteq \mathcal{L}_{(S,\nu)}(B)$, thus
  $\mathcal{L}_{(P,\nu)}(\overline{B}) \subseteq
  \mathcal{L}_{(S,\nu)}(\overline{B})$. We obtain that
  $\mathcal{L}_{(\vec{q},P,\nu)}(\mathcal{A}^e \times \overline{B})
  \subseteq \mathcal{L}_{(\vec{r},\nu)}(\mathcal{A}^e) \cap
  \mathcal{L}_{(S,\nu)}(\overline{B}) =
  \mathcal{L}_{(\vec{r},S,\nu)}(\mathcal{A}^e \times \overline{B})$.
  Since moreover $\Phi \rightarrow \Psi$, we have that
  $\mathcal{L}_{(\vec{q},P,\Phi)}(\mathcal{A}^e \times \overline{B})
  \subseteq \mathcal{L}_{(\vec{r},S,\Phi)}(\mathcal{A}^e \times
  \overline{B}) \subseteq \mathcal{L}_{(\vec{r},S,\Psi)}(\mathcal{A}^e
  \times \overline{B})$. \qed}

\begin{example} In the antichain from Fig. \ref{fig:refinement} (d), 
$(\tuple{q_1^1,q_1^2},\set{p_1},v=1) \sqsubseteq_{\mathit{img}}
  (\tuple{q_1^1,q_1^2},\set{p_1},$ $v=1)$ because $\tuple{q_1^1,q_1^2} =
  \tuple{q_1^1,q_1^2}$, $\set{p_1} \supseteq \set{p_1}$, and $v=1
  \rightarrow v=1$. $\blacksquare$ \end{example}

As a remark, the language inclusion algorithm for non-deterministic
automata on finite alphabets \cite{abdulla} uses also a more
sophisticated subsumption relation based on a pre-computed simulation
\cite{milner} between the states of the automata. We have defined
a~similar notion of simulation for data automata and an algorithm for
computing such simulations \cite{tech-report}. The integration of data
simulations within the framework of antichain-based abstraction
refinement and its practical assessment are considered as future work.

\subsection{The Trace Inclusion Semi-algorithm}
\label{sec:trace-inclusion-algorithm}

With the previous definitions, Algorithm~\ref{alg:trace-inclusion} describes the
procedure for checking trace inclusion. It uses a classical worklist iteration
loop (lines \ref{ln:beginWhile}-\ref{ln:endWhile}) that builds an antichain tree
by simultaneously unfolding the expansion $\mathcal{A}^e$ of the network
$\mathcal{A}$ and the complement $\overline{B}$ of the the observer $B$, while
searching for a counterexample trace $w \in \mathcal{L}(\mathcal{A}^e \times
\overline{B})$. Both $\mathcal{A}^e$ and $\overline{B}$ are built on-the-fly,
during the abstract state space exploration. 

\begin{algorithm}[t!]
{\scriptsize\begin{algorithmic}[0]
  \State {\bf input}: 
  \begin{compactenum}
    \item a DAN $\mathcal{A} = \tuple{A_1,\ldots,A_N}$ such that $A_i
      =
      \tuple{\mathcal{D},\Sigma_i,\vec{x}_i,Q_i,\iota_i,F_i,\prod_i}$
      for all $i \in [1,N]$,
    \item a DA $B =
      \tuple{\mathcal{D},\Sigma,\vec{x}_B,Q_B,\iota_B,F_B,\prod_B}$
      such that $\vec{x}_B \subseteq \bigcup_{i=1}^N\vec{x}_i$ .
  \end{compactenum}
  \State {\bf output}: true if $\lang{\mathcal{A}}\proj_{\vec{x}_B}
    \subseteq \lang{B}$, otherwise a trace $\tau \in
    \lang{\mathcal{A}}\proj_{\vec{x}_B} \setminus \lang{B}$ .
\end{algorithmic}

\begin{algorithmic}[1]
  \State  $\Pi \leftarrow \emptyset$, $\mathtt{Visited} \leftarrow \emptyset$, 
  $\mathtt{Next} \leftarrow \tuple{\mathit{root}_{\tuple{\mathcal{A},B}}, \varepsilon}$, 
  $\mathtt{Subsume} \leftarrow \emptyset$\label{ln:init}

  \While {$\mathtt{Next} \neq \emptyset$}\label{ln:beginWhile}

  \State chose $\mathtt{curr} \in \mathtt{Next}$ and move
  $\mathtt{curr}$ from $\mathtt{Next}$ to
  $\mathtt{Visited}$\label{ln:move}

  \State match $\mathtt{curr}$ with $\tuple{s,p}$

  \If{$s$ is an accepting product state}\label{ln:ifAccepting}

  \State let $\rho$ be the path from the root to $\mathtt{curr}$ and $k$ be the pivot of $\rho$\label{ln:pivot}

  \If{$k\geq0$}\label{ln:spuriousnessCheck}

  \State $\Pi \leftarrow \Call{refinePredicateMapByInterpolation}{\Pi,\rho,k}$\label{ln:updateMap}

  \State $\mathtt{rem} \leftarrow \Call{subTree}{\rho_k}$\label{ln:subTree}

  \For{$(n,m) \in \mathtt{Subsume} ~\mbox{such that}~ m \in \mathtt{rem}$}\label{ln:forSubRem}

  \State move $n$ from $\mathtt{Visited}$ to $\mathtt{Next}$\label{ln:moveNext}

  \EndFor

  \State remove $\mathtt{rem}$ from $(\mathtt{Visited},\mathtt{Next},\mathtt{Subsume})$\label{ln:removeSubTree}

  \State add $\rho_k$ to $\mathtt{Next}$\label{ln:addPivot}

  \Else 
  
  \State return $\Call{extractCounterexample}{\rho}$\label{ln:realCEXreport}

  \EndIf 

  \Else

  \State $i \leftarrow 0$

  \For{$t \in \mathit{Post}_\Pi(s)$}\label{ln:post}

  \If{there exists $m = \tuple{t',p'} \in \mathtt{Visited} \text{ such that } t \sqsubseteq t'$}

  \State add $(\mathtt{curr},m)$ to $\mathtt{Subsume}$\label{ln:addSubsume}

  \Else

  \State $\mathtt{rem} \leftarrow \set{n \in \mathtt{Next} \mid n=\tuple{t',p'} \text{ and } t' \sqsubset t}$\label{ln:rem}

  \State $\mathtt{succ} \leftarrow \tuple{t,p.i}$\label{ln:buildSucc}

  \State $i \leftarrow i+1$

  \For{$n \in \mathtt{Visited} \text{ such that } n \text{ has a successor } m\in\mathtt{rem}$}

  \State add $(n,\mathtt{succ})$ to $\mathtt{Subsume}$\label{ln:addSuccSubsume}

  \EndFor

  \For{$(n,m) \in \mathtt{Subsume} \text{ such that } m \in \mathtt{rem}$}
  
  \State add $(n,\mathtt{succ})$ to $\mathtt{Subsume}$\label{ln:addSubSubsume}

  \EndFor

  \State remove $\mathtt{rem}$ from $(\mathtt{Visited}, \mathtt{Next}, \mathtt{Subsume})$\label{ln:removeRem}

  \State add $\mathtt{succ}$ to $\mathtt{Next}$\label{ln:addNext}

  \EndIf

  \EndFor

  \EndIf

  \EndWhile\label{ln:endWhile}
\end{algorithmic}}
\caption{Trace Inclusion Semi-algorithm}\label{alg:trace-inclusion}
\end{algorithm}

The processed antichain nodes are kept in the set $\mathtt{Visited}$,
and their abstract successors, not yet processed, are kept in the set
$\mathtt{Next}$. Initially, $\mathtt{Visited}=\emptyset$ and
$\mathtt{Next}=\set{\mathit{root}_{\mathcal{A},B}}$. The algorithm
uses a predicate map $\Pi$, which is initially empty (line
\ref{ln:init}).


We keep a set of subsumption edges $\mathtt{Subsume} \subseteq
\mathtt{Visited} \times \left(\mathtt{Visited} \cup
\mathtt{Next}\right)$ with the following meaning:
$(\tuple{s,p},\tuple{t,q}) \in \mathtt{Subsume}$ for two antichain
nodes, where $s,t$ are product states and $p,q \in \nat^*$ are tree
positions, if and only if there exists an abstract successor $s' \in
\mathit{Post}_\Pi(s)$ such that $s' \sqsubseteq t$ (Definition
\ref{def:subsumption}). Observe that we do not explicitly store a
subsumed successor of a product state $s$ from the antichain; instead,
we add a~subsumption edge between the node labeled with $s$ and the
node that subsumes that particular successor. The algorithm terminates
when each abstract successors of a~node from $\mathtt{Next}$ is
subsumed by some node from $\mathtt{Visited}$.


An iteration of Algorithm \ref{alg:trace-inclusion} starts by chosing a current
antichain node $\mathtt{curr}=\tuple{s,p}$ from $\mathtt{Next}$ and moving it to
$\mathtt{Visited}$ (line~\ref{ln:move}). If the product state $s$ is accepting
(line~\ref{ln:ifAccepting}) we check the counterexample path $\rho$, from the
root of the antichain to $\mathtt{curr}$, for spuriousness, by computing its
pivot $k$. If $k\geq0$, then $\rho$ is a~spurious counterexample
(line~\ref{ln:spuriousnessCheck}), and the path formula of the suffix of $\rho$,
which starts with position $k$, is infeasible. In this case, we compute an
interpolant for the suffix and refine the current predicate map $\Pi$ by adding
the predicates from the interpolant to the corresponding substates of the
product states from the suffix (line \ref{ln:updateMap}).

The computation of the interpolant and the update of the predicate map are
done by the $\Call{refinePredicateMapByInterpolation}{}$ function using the
approach described in Section \ref{sec:cex}. Subsequently, we remove (line
\ref{ln:removeSubTree}) from the current antichain the subtree rooted at the
pivot node $\rho_k$, i.e.\ the $k$-th node on the path $\rho$
(line~\ref{ln:subTree}),~and~add $\rho_k$ to $\mathtt{Next}$ in order to trigger
a recomputation of this subtree with the new predicate map. Moreover, all nodes
with a successor previously subsumed by a node in the removed subtree are moved
from $\mathtt{Visited}$ back to $\mathtt{Next}$ in order to reprocess
them~(line~\ref{ln:moveNext}).

On the other hand, if the counterexample $\rho$ is found to be real
($k=-1$), any valuation $\nu \in
\bigcup_{i=0}^{\len{\rho}}\mathcal{D}^{\vec{x}_{\mathcal{A}}^i}$ that
satisfies the path formula $\Theta(\rho)$ yields a counterexample
trace $w \in \mathcal{L}(\mathcal{A})\proj_{\vec{x}_B} \setminus
\mathcal{L}(B)$, obtained by ignoring all variables from
$\vec{x}_{\mathcal{A}} \setminus \vec{x}_B$ (line
\ref{ln:realCEXreport}).

If the current node is not accepting, we generate its abstract
successors (line \ref{ln:post}). In order to keep in the antichain
only nodes that are incomparable w.r.t. the subsumption relation
$\sqsubseteq$, we add a successor $t$ of $s$ to $\mathtt{Next}$ (lines
\ref{ln:buildSucc} and \ref{ln:addNext}) only if it is not subsumed by
another product state from a node $m\in\mathtt{Visited}$. Otherwise,
we add a subsumption edge $(\mathtt{curr},m)$ to the set
$\mathtt{Subsume}$ (line \ref{ln:addSubsume}). Furthermore, if $t$ is
not subsumed by another state in $\mathtt{Visited}$, we remove from
$\mathtt{Next}$ all nodes $\tuple{t',p'}$ such that $t$ strictly
subsumes $t'$ (lines \ref{ln:rem} and \ref{ln:removeRem}) and add
subsumption edges to the node storing $t$ from all nodes with a
removed successor (line \ref{ln:addSuccSubsume}) or a removed
subsumption edge (line \ref{ln:addSubSubsume}).

The following theorem states the soundness of our trace inclusion
semi-algorithm. The theorem is proved in the appendix together with
the other results presented above.

\begin{theorem}\label{thm:trace-inclusion-soundness}
  Let $\mathcal{A} = \tuple{A_1,\ldots,A_N}$ be a DAN such that $A_i
  = \tuple{\mathcal{D},\Sigma_i,\vec{x}_i,Q_i,\iota_i,F_i,\prod_i}$
  for all $i \in [1,N]$, and let $B =
  \tuple{\mathcal{D},\Sigma,\vec{x}_B,Q_B,\iota_B,F_B,\prod_B}$ be a
  DA such that $\vec{x}_B \subseteq \bigcup_{i=1}^N\vec{x}_i$. If
  Algorithm \ref{alg:trace-inclusion} terminates and returns true on
  input $\mathcal{A}$ and $B$, then
  $\mathcal{L}(\mathcal{A})\proj_{\vec{x}_B} \subseteq
  \mathcal{L}(B)$. 
\end{theorem}

The dual question ``if there exists a counterexample trace $w \in
\mathcal{L}(\mathcal{A})\proj_{\vec{x}_B} \setminus \mathcal{L}(B)$,
will Algorithm~\ref{alg:trace-inclusion} discover it?'' can also be
answered positively, using an implementation that enumerates the
abstract paths in a systematic way, e.g.\ by using a~breadth-first
path exploration. This can be done using a queue to implement the
$\mathtt{Next}$ set in Algorithm \ref{alg:trace-inclusion}.

\subsection{Proof of Theorem \ref{thm:trace-inclusion-soundness}}

Given a network $\mathcal{A}=\tuple{A_1,\ldots,A_N}$ where $A_i =
\tuple{\mathcal{D},\Sigma_i,\vec{x}_i,Q_i,\iota_i,F_i,\prod_i}$ for
all $i \in [1,N]$ and an observer $B =
\tuple{\mathcal{D},\Sigma,\vec{x}_B,Q_B,\iota_B,F_B,\prod_B}$, we
recall that a configuration of the product automaton $\mathcal{A}^e
\times \overline{B}$ is a tuple $(\tuple{q_1,\ldots,q_N},P,\nu) \in
Q_1 \times \ldots \times Q_N \times 2^{Q_B} \times
\mathcal{D}^{\vec{x}_{\mathcal{A}}}$, and a node of the antichain
$\mathcal{T}$ is a pair $\tuple{s,p}$ where $s$ is a product state for
$\mathcal{A}$ and $B$ and $p\in\nat^*$ is a tree position. Moreover,
$\mathit{root}_{\tuple{\mathcal{A},B}} =
(\tuple{\iota_1,\ldots,\iota_N}, \set{\iota_B}, \true)$ is the product
state that labels the root of $\mathcal{T}$.  In the following, let
$\Gamma = (\Pi,\mathtt{Visited},\mathtt{Next},\mathtt{Subsume})$ be an
\emph{antichain state} where $\Pi$ is the predicate map, and
$\mathtt{Visited}$, $\mathtt{Next}$, and $\mathtt{Subsume}$ are the
sets of antichain nodes handled by Algorithm
\ref{alg:trace-inclusion}.

We say that $\Gamma$ is a \emph{closed
  antichain state} if and only if, for all nodes $\tuple{s,p} \in
\mathtt{Visited}$ and every successor $(\vec{q},P,\nu) \in
\mathit{succ}_{\mathcal{A}^e \times \overline{B}}(\sem{s})$ of 
a~configuration of the product automaton $\mathcal{A}^e \times
\overline{B}$ represented by the product state $s$, there exists a
node $\tuple{t,r} \in \mathtt{Visited} \cup \mathtt{Next}$ such that
$\mathcal{L}_{(\vec{q},P,\nu)}(\mathcal{A}^e \times \overline{B})
\subseteq \mathcal{L}_t(\mathcal{A}^e \times \overline{B})$ and one of
the following holds: \begin{compactitem}
  \item $r = p.i$ for some $i \in \nat$, i.e.\ $\tuple{t,r}$ is a
    child of $\tuple{s,p}$ in the antichain
    $\mathcal{T}=\mathtt{Visited}\cup\mathtt{Next}$, or
  \item $(\tuple{s,p}, \tuple{t,r}) \in \mathtt{Subsume}$.
  \end{compactitem}
In other words, the current antichain $\mathcal{T}$, defined as the
union of the sets $\mathtt{Visited}$ and $\mathtt{Next}$, is in a
closed state, if the residual language of every successor of a
configuration of the product automaton $\mathcal{A}^e \times
\overline{B}$ that is covered by a visited product state must be
included in the residual language of a product state stored in the
antichain, either as a direct successor in the tree or via a
subsumption edge.

For a product state $s$, we define $\mathit{Dist}(s) =
\min\set{\len{w} \mid w \in \mathcal{L}_s(\mathcal{A}^e \times
  \overline{B})}$, and $\mathit{Dist}(s)=\infty$ if and only if
$\mathcal{L}_s(\mathcal{A}^e \times \overline{B}) = \emptyset$. For a
finite non-empty set of antichain nodes $S$, we define
$\mathit{Dist}(S) = \min\set{\mathit{Dist}(s) \mid \tuple{s,p} \in
  S}$ with $\mathit{Dist}(\emptyset) = \infty$. 

\begin{lemma}\label{lemma:succ-sem}
  Given a network $\mathcal{A}$ and an observer $B$, for any product
  state $s$ of $\mathcal{A}$ and $B$, we have
  $\mathit{succ}_{\mathcal{A}^e \times \overline{B}}(\sem{s}) =
  \bigcup_{t \in \mathit{Post}(s)} \sem{t}$.
\end{lemma}
\proof{ Let $s = (\vec{q},P,\Phi)$. ``$\subseteq$'' Let
  $(\vec{r},S,\mu) \in \mathit{succ}_{\mathcal{A}^e \times
    \overline{B}}(\sem{s})$ be a configuration of $\mathcal{A}^e
  \times \overline{B}$ for which there exists $(\vec{q},P,\nu) \in
  \sem{s}$ such that $(\vec{q},P,\nu) \arrow{\sigma,\theta}{}
  (\vec{r},S,\mu)$. Then there exists a unique rule $(\vec{q},P)
  \arrow{\sigma,\theta}{} (\vec{r},S) \in \Delta^p$ such that
  $(\nu,\mu) \modelsthd \theta$. Moreover, if $(\vec{q},P,\nu) \in
  \sem{s}$, we have $\nu \modelsthd \Phi$. Let $t = (\vec{r},S,\Psi)
  \in \mathit{Post}(s)$ where $\Psi(\vec{x}_{\mathcal{A}}) \equiv
  \exists \vec{x}'_{\mathcal{A}} ~.~ \Phi(\vec{x}'_{\mathcal{A}})
  \wedge \theta(\vec{x}'_{\mathcal{A}}, \vec{x}_{\mathcal{A}})$. We
  have $\mu \modelsthd \Psi$, hence $(\vec{r},S,\mu) \in
  \sem{t}$. ``$\supseteq$'' Let $(\vec{r},S,\mu) \in \sem{t}$ for some
  $t \in \mathit{Post}(s)$. Then we have $t = (\vec{r},S,\Psi)$ where
  $\Psi(\vec{x}_{\mathcal{A}}) \equiv \exists \vec{x}'_{\mathcal{A}}
  ~.~ \Phi(\vec{x}'_{\mathcal{A}}) \wedge
  \theta(\vec{x}'_{\mathcal{A}}, \vec{x}_{\mathcal{A}})$. Since $\mu
  \modelsthd \Psi$, there exists $\nu \modelsthd \Phi$ such that
  $(\vec{q},P,\nu) \arrow{\sigma,\theta}{} (\vec{r},S,\mu)$. Hence
  $(\vec{q},P,\nu) \in \sem{s}$, thus $(\vec{r},S,\mu) \in
  \mathit{succ}_{\mathcal{A}^e \times \overline{B}}(\sem{s})$. \qed}\bigskip
 
\begin{lemma}\label{lemma:post-abs}
  Given a network $\mathcal{A}$, an observer $B$, and a predicate map
  $\Pi$, for any product state $s$ of $\mathcal{A}^e \times \overline{B}$ and any
  product state $t \in \mathit{Post}(s)$ there exists $t' \in
  \mathit{Post}_\Pi(s)$ such that $\sem{t} \subseteq \sem{t'}$.
\end{lemma}
\proof{ Let $t = (\vec{r},S,\Psi) \in \mathit{Post}(s)$. By the
  definition of $\mathit{Post}_\Pi$, we have
  $t'=(\vec{r},S,\Psi^\sharp) \in \mathit{Post}_\Pi(s)$, where $\Psi
  \rightarrow \Psi^\sharp$, thus $\sem{t} \subseteq \sem{t'}$. \qed}\bigskip

\begin{lemma}\label{lemma:succ-post}
  Given a network $\mathcal{A}$, an observer $B$, and a predicate map
  $\Pi$, for each product state $s$ and each configuration
  $(\vec{q},P,\nu) \in \mathit{succ}_{\mathcal{A}^e \times
    \overline{B}}(\sem{s})$ there exists a product state $t \in
  \mathit{Post}_\Pi(s)$ such that $(\vec{q},P,\nu) \in \sem{t}$.
\end{lemma}
\proof{ We use the fact that $\mathit{succ}_{\mathcal{A}^e \times
    \overline{B}}(\sem{s}) = \bigcup_{t \in \mathit{Post}(s)} \sem{t}$
  (Lemma \ref{lemma:succ-sem}) and that for each $t \in
  \mathit{Post}(s)$ there exists $t' \in \mathit{Post}_\Pi(s)$ such
  that $\sem{t} \subseteq \sem{t'}$ (Lemma
  \ref{lemma:post-abs}). \qed}\bigskip

The proof of soundness of Algorithm \ref{alg:trace-inclusion} relies
on the inductive invariants ($\mathit{Inv}_1$) and ($\mathit{Inv}_2$)
from the following lemma.

\begin{lemma}\label{lemma:invariants}
  The following invariants hold each time line \ref{ln:beginWhile} is
  reached in Algorithm \ref{alg:trace-inclusion}:
  \begin{compactitem}
  \item ($\mathit{Inv}_1$)
    $\Gamma=(\Pi,\mathtt{Visited},\mathtt{Next},\mathtt{Subsume})$
    is closed,
  \item ($\mathit{Inv}_2$)
    $\mathit{Dist}(\mathit{root}_{\tuple{\mathcal{A},B}}) < \infty
    \Rightarrow \mathit{Dist}(\mathtt{Visited}) >
    \mathit{Dist}(\mathtt{Next})$.
  \end{compactitem}
\end{lemma}
\proof{Initially, when coming to line \ref{ln:beginWhile} for the first time,
  we have $\mathtt{Visited}=\emptyset$, thus
  $\mathit{Dist}(\mathtt{Visited}) = \infty$, and both invariants hold
  trivially. For the case when coming to line \ref{ln:beginWhile} after
  executing the loop body, we denote by:
  \[\begin{array}{rcl}
  \Gamma_\old & = & (\Pi_\old,\mathtt{Visited}_\old,\mathtt{Next}_\old,\mathtt{Subsume}_\old) \\
  \Gamma_\new & = & (\Pi_\new,\mathtt{Visited}_\new,\mathtt{Next}_\new,\mathtt{Subsume}_\new)
  \end{array}\]
  the antichain states before and after the execution of the main
  loop. We assume that both invariants hold for $\Gamma_\old$.

  \vspace*{\baselineskip}\noindent ($\mathit{Inv}_1$) Let $\tuple{s,p}
  \in \mathtt{Visited}_\new$ and $(\vec{q},P,\nu) \in
  \mathit{succ}_{\mathcal{A}^e \times \overline{B}}(\sem{s})$. We
  distinguish two cases according to the control path taken inside
  the main loop:
  \begin{compactenum}[(1)]
  \item If the test on line \ref{ln:ifAccepting} is positive, the
    predicate map is augmented, i.e.\ $\Pi_\new \supseteq \Pi_\old$
    (line \ref{ln:updateMap}). Let $\Gamma' = (\Pi_\new,
    \mathtt{Visited}_\old, \mathtt{Next}_\old, \mathtt{Subsume}_\old)$
    be the next antichain state. Clearly $\Gamma'$ is closed provided
    that $\Gamma_\old$ is. Next, let $n_{\mathit{pivot}} \in
    \mathtt{Visited}_\old$ be the pivot of the path to the current
    node (line \ref{ln:pivot}) and define the following sets of nodes:
    \[\begin{array}{rcl}
    T & = & \Call{subTree}{n_{\mathit{pivot}}} \\
    S & = & \set{n \in \mathtt{Visited}_\old \mid \exists m \in T ~.~ (n,m) \in \mathtt{Subsume}_\old}
    \end{array}\]
    Then we obtain (lines \ref{ln:forSubRem}--\ref{ln:addPivot}):
    \[\begin{array}{rcl}
    \mathtt{Visited}_\new & = & \mathtt{Visited}_\old \setminus (S \cup T) \\
    \mathtt{Next}_\new & = & ((\mathtt{Next}_\old \cup S) \setminus T) \cup \set{n_{\mathit{pivot}}} \\
    \mathtt{Visited}_\new \cup \mathtt{Next}_\new & = & ((\mathtt{Visited}_\old \cup \mathtt{Next}_\old) \setminus T) 
    \cup \set{n_{\mathit{pivot}}}
    \end{array}\]
    Since $\Gamma'$ is closed, there exists a node $\tuple{t,r} \in
    \mathtt{Visited}_\old\cup\mathtt{Next}_\old$ such that
    $\mathcal{L}_{(\vec{q},P,\nu)}(\mathcal{A}^e \times \overline{B})
    \subseteq \mathcal{L}_t(\mathcal{A}^e \times \overline{B})$ and
    either $r=p.i$ for some $i \in \nat$ or
    $(\tuple{s,p},\tuple{t,r})\in\mathtt{Subsume}_\old$. We
    distinguish two cases:
    \begin{compactenum}[(a)]
    \item $\tuple{t,r} \not\in T$. Then $\tuple{t,r} \in
      \mathtt{Visited}_\new \cup \mathtt{Next}_\new$ and, because
      $\mathtt{Subsume}_\new = \mathtt{Subsume}_\old \cap
      (\mathtt{Visited}_\new \times (\mathtt{Visited}_\new \cup
      \mathtt{Next}_\new))$, we obtain that $\Gamma_\new$ is closed as
      well.
    \item $\tuple{t,r} \in T$. Then we distinguish two further cases:
      \begin{compactenum}[(i)]
      \item If $r=p.i$ for some $i\in\nat$, since we have assumed that
        $\tuple{s,p} \in \mathtt{Visited}_\new$, we have $\tuple{s,p}
        \not\in T$. The only possibility is then
        $\tuple{t,r}=n_{\mathit{pivot}}$ and $\tuple{s,p}$ is the
        parent of $n_{\mathit{pivot}}$. In this case, we have
        $\tuple{t,r} \in \mathtt{Next}_\new$. 
      \item If $(\tuple{s,p},\tuple{t,r}) \in \mathtt{Subsume}_\old$,
        then $\tuple{s,p} \in S$, which contradicts the assumption
        $\tuple{s,p} \in \mathtt{Visited}_\new$. 
      \end{compactenum}
    \end{compactenum}
  \item Otherwise, the test on line \ref{ln:ifAccepting} is negative,
    in which case we have $\Pi_\new = \Pi_\old$ and
    $\mathtt{Visited}_\new = \mathtt{Visited}_\old \cup
    \set{\mathtt{curr}}$. For each $(\vec{q},P,\nu) \in
    \mathit{succ}_{\mathcal{A}^e \times \overline{B}}(\sem{s})$ there
    exists $t \in \mathit{Post}_\Pi(s)$ such that
    $\mathcal{L}_{(\vec{q},P,\nu)}(\mathcal{A}^e \times \overline{B})
    \subseteq \mathcal{L}_t(\mathcal{A}^e \times \overline{B})$ (by
    Lemma \ref{lemma:succ-post}). We distinguish two cases:
    \begin{compactenum}[(a)]
    \item $\tuple{s,p}=\mathtt{curr}$. In this case, either\begin{inparaenum}[(i)]%
      \item there is $\tuple{t',p'}\in\mathtt{Visited}_\old$ such
        that $t \sqsubseteq t'$, and then we also have
        $\mathcal{L}_{(\vec{q},P,\nu)}(\mathcal{A}^e \times
        \overline{B}) \subseteq \mathcal{L}_{t'}(\mathcal{A}^e \times
        \overline{B})$ (Definition \ref{def:subsumption}) and
        $(\tuple{s,p},\tuple{t',p'}) \in \mathtt{Subsume}_\new$ (added
        on line \ref{ln:addSubsume}), or
      \item $(t,p.i) \in \mathtt{Next}_\new$ for some $i \in \nat$
        (added on lines \ref{ln:buildSucc} and \ref{ln:addNext}).
        \end{inparaenum}
    \item Otherwise $\tuple{s,p} \in \mathtt{Visited}_\old$. As
      $\Gamma'$ is closed, there is $\tuple{u,r} \in
      \mathtt{Visited}_\old \cup \mathtt{Next}_\old$ such that
      $\mathcal{L}_{(\vec{q},P,\nu)}(\mathcal{A}^e \times
      \overline{B}) \subseteq \mathcal{L}_u(\mathcal{A}^e \times
      \overline{B})$ and either $r=p.i$ for some $i\in\nat$ or
      $(\tuple{s,p},\tuple{u,r}) \in \mathtt{Subsume}_\old$.  We
      distinguish two sub-cases: \begin{compactenum}[(i)]
        \item $\tuple{u,r} \in \mathtt{rem}$ (line \ref{ln:rem}).
          Then $\mathcal{L}_u(\mathcal{A}^e \times \overline{B})
          \subseteq \mathcal{L}_t(\mathcal{A}^e \times \overline{B})$
          (Definition \ref{def:subsumption}), hence
          $\mathcal{L}_{(\vec{q},P,\nu)}(\mathcal{A}^e \times
          \overline{B}) \subseteq \mathcal{L}_t(\mathcal{A}^e \times
          \overline{B})$. If $r=p.i$, then
          $(\tuple{s,p},\tuple{t,r'}) \in \mathtt{Subsume}_\new$ for
          some $r'\in\nat^*$ (added on line
          \ref{ln:addSuccSubsume}). Else, if
          $(\tuple{s,p},\tuple{u,r}) \in \mathtt{Subsume}_\old$, we
          have $(\tuple{s,p},\tuple{t,r'}) \in \mathtt{Subsume}_\new$
          for some $r'\in\nat^*$ (added on line
          \ref{ln:addSubSubsume}). In both cases, we obtain that
          $\Gamma_\new$ is closed.
        \item $\tuple{u,r} \not\in \mathtt{rem}$. Then
          $\tuple{u,r} \in \mathtt{Visited}_\new \cup
          \mathtt{Next}_\new$. Since $\mathtt{Subsume}_\new =
          \mathtt{Subsume}_\old \cap (\mathtt{Visited}_\new \times
          (\mathtt{Visited}_\new \cup \mathtt{Next}_\new))$, we obtain
          that $\Gamma_\new$ is closed.
        \end{compactenum}
    \end{compactenum}
  \end{compactenum}

  \vspace*{\baselineskip}\noindent ($\mathit{Inv}_2$) We distinguish two cases:
  \begin{compactenum}
  \item If $\mathit{Dist}(\mathtt{Visited}_\new) = \infty$, it is
    sufficient to show that $\mathit{Dist}(\mathtt{Next}_\new) <
    \infty$. Suppose, by contradiction, that
    $\mathit{Dist}(\mathtt{Next}_\new) = \infty$, hence
    $\mathit{Dist}(\mathtt{Visited}_\new \cup \mathtt{Next}_\new) =
    \infty$, and since $\mathit{root}_{\tuple{\mathcal{A},B}} \in
    \mathtt{Visited}_\new \cup \mathtt{Next}_\new$, we obtain
    $\mathit{Dist}(\mathit{root}_{\tuple{\mathcal{A},B}}) = \infty$,
    contradiction. 
  \item Otherwise, $\mathit{Dist}(\mathtt{Visited}_\new) < \infty$ and
    there exists a node $\tuple{s,p} \in \mathtt{Visited}_\new$ such
    that $\mathit{Dist}(\mathtt{Visited}_\new) = \mathit{Dist}(s) <
    \infty$. Let $w=(\nu_0,\sigma_0), (\nu_1,\sigma_1), \ldots,
    (\nu_n,\diamond) \in \mathcal{L}_s(\mathcal{A}^e \times
    \overline{B})$ be a trace such that
    $\mathit{Dist}(\mathtt{Visited}_\new) = n$. Then there exists a
    run $(\vec{q}_0,P_0,\nu_0) \arrow{\sigma_0}{}
    (\vec{q}_1,P_1,\nu_1) \arrow{\sigma_1}{} \ldots
    \arrow{\sigma_{n-1}}{} (\vec{q}_n,P_n,\nu_n)$ of $\mathcal{A}^e
    \times \overline{B}$ over $w$ such that $(\vec{q}_0,P_0,\nu_0) \in
    \sem{s}$ and $(\vec{q}_n,P_n)$ a final state of $\mathcal{A}^e
    \times \overline{B}$. Since $\Gamma_\new$ is closed due to
    ($\mathit{Inv}_1$) and $(\vec{q}_1,P_1,\nu_1) \in
    \mathit{succ}_{\mathcal{A}^e \times \overline{B}}(\sem{s})$, there
    exists a node $\tuple{s_1,p_1} \in \mathtt{Visited}_\new \cup
    \mathtt{Next}_\new$ such that
    $\mathcal{L}_{(\vec{q}_1,P_1,\nu_1)}(\mathcal{A}^e \times
    \overline{B}) \subseteq \mathcal{L}_{s_1}(\mathcal{A}^e \times
    \overline{B})$. If $\tuple{s_1,p_1} \in \mathtt{Next}_\new$, we
    obtain that $\mathit{Dist}(\mathtt{Next}_\new) \leq n-1$, and we
    are done. Otherwise, $\tuple{s_1,p_1} \in \mathtt{Visited}_\new$,
    and we can repeat the same argument inductively, to discover a
    sequence of nodes $\tuple{s_1,p_1}, \ldots, \tuple{s_n,p_n} \in
    \mathtt{Visited}_\new$ such that
    $\mathcal{L}_{(\vec{q}_i,P_i,\nu_i)}(\mathcal{A}^e \times
    \overline{B}) \subseteq \mathcal{L}_{s_n}(\mathcal{A}^e \times
    \overline{B})$ for all $i \in [1,n]$. Since $(\vec{q}_n,P_n)$ is a
    final state of $\mathcal{A}^e \times \overline{B}$, we have
    $(\nu_n,\diamond) \in
    \mathcal{L}_{(\vec{q}_i,P_i,\nu_i)}(\mathcal{A}^e \times
    \overline{B})$, thus $(\nu_n,\diamond) \in
    \mathcal{L}_{s_n}(\mathcal{A}^e \times \overline{B})$, and $s_n$
    is an accepting product state. But this contradicts with the fact
    that accepting product states are never stored in the antichain.
  \end{compactenum}
\qed}\bigskip

Back to the proof of Theorem \ref{thm:trace-inclusion-soundness}:

\proof{ If Algorithm \ref{alg:trace-inclusion} terminates and reports
  true, this is because $\mathtt{Next}=\emptyset$, hence
  $\mathit{Dist}(\mathtt{Next})=\infty$. By Lemma
  \ref{lemma:invariants} ($\mathit{Inv}_2$), we obtain that
  $\mathit{Dist}(\mathit{root}_{\tuple{\mathcal{A},B}})=\infty$.
  Suppose, by contradiction, that
  $\mathcal{L}(\mathcal{A})\proj_{\vec{x}_B} \not\subseteq
  \mathcal{L}(B)$. By Lemma \ref{lemma:inclusion-emptiness}, there
  exists a trace \[w = (\nu_0,\sigma_0)(\nu_1,\sigma_1) \ldots
  (\nu_n,\diamond) \in \mathcal{L}(\mathcal{A}^e \times
  \overline{B})\enspace.\] Thus we have a run of $\mathcal{A}^e \times
  \overline{B}$ over $w$:
  \[(\vec{q}_0,P_0,\nu_0) \arrow{\sigma_0}{} (\vec{q}_1,P_1,\nu_1) \arrow{\sigma_1}{} 
  \ldots \arrow{\sigma_{n-1}}{} (\vec{q}_n,P_n,\nu_n)\] where
  $\vec{q}_0 = \tuple{\iota_1,\ldots,\iota_N}$, $P_0=\set{\iota_B}$,
  $\vec{q}_n$ is final in $\mathcal{A}^e$, $P_n \cap F_B =
  \emptyset$. But, since $(\vec{q}_0,P_0,\nu_0) \in
  \sem{\mathit{root}_{\tuple{\mathcal{A},B}}}$, we have $w \in
  \mathcal{L}_{\mathit{root}_{\tuple{\mathcal{A},B}}}(\mathcal{A}^e
  \times \overline{B})$. Hence,
  $\mathit{Dist}(\mathit{root}_{\tuple{\mathcal{A},B}}) \leq n$, which
  is in contradiction with the fact that
  $\mathit{Dist}(\mathit{root}_{\tuple{\mathcal{A},B}})=\infty$. Consequently,
  it must be the case that $\mathcal{L}(\mathcal{A})\proj_{\vec{x}_B}
  \subseteq \mathcal{L}(B)$. \qed}

\section{Computing Simulations of Data Automata}
\label{sec:simulations}

In the case of classical Rabin-Scott finite automata over finite
alphabets, a \emph{simulation} \cite{milner} is a relation
on the states of an automaton, which is invariant with respect to the
transition relation of the automaton. The simulation-based approach
for checking language inclusion between two automata $A$ and $B$ first
computes a simulation relation on the union of the states of $A$ and
$B$, then checks whether the pair of initial states is a member of the
simulation relation. This method is not complete because there exist
automata, such that $\mathcal{L}(A) \subseteq \mathcal{L}(B)$, but the
initial state of $A$ is not simulated by the initial state of $B$. A
pre-computed simulation relation can be used however to speed up the
convergence of the antichain-based method, by weakening the
subsumption relation used by the antichain construction algorithm
\cite{abdulla}.

In this section, we define a notion of simulation between data
automata and give an algorithm that computes useful
under-approximations of the weakest simulation on a data
automaton. The simulation relation can be used to enhance the
convergence of Algorithm \ref{alg:trace-inclusion}, similar to the way
in which classical simulations are integrated with the antichain-based
language inclusion algorithm for automata over finite alphabets
\cite{abdulla}.

\begin{definition}\label{def:data-simulation}
  A relation $R \subseteq Q \times \mathcal{D}^{\vec{x}} \times Q$ is
  a \emph{data simulation} for a DA $A = \tuple{\Sigma, \mathcal{D},
    \vec{x}, Q, \iota, F, \prod}$ if and only if, for all $(q,\nu,q')
  \in R$ the following hold: \begin{compactenum}
  \item\label{it1:data-simulation} $q \in F$ only if $q' \in F$, and
  \item\label{it2:data-simulation} for all $\sigma\in\Sigma$ and all
    $(r,\nu') \in Q \times \mathcal{D}^{\vec{x}}$ such that $(q,\nu)
    \arrow{\sigma}{}_A~ (r,\nu')$ there exists $r' \in Q$ such that
    $(q',\nu) \arrow{\sigma}{}_A~ (r',\nu')$ and $(r,\nu',r') \in R$.
  \end{compactenum}
\end{definition}
Observe that, while a classical simulation is a binary relation on
states, a data simulation is a ternary relation involving also a
valuation of the variables. The following lemma shows that a data
simulation preserves residual languages.

\begin{lemma}\label{lemma:simulation-residual}
  Given a DA $A=\tuple{\Sigma, \mathcal{D}, \vec{x}, Q, \iota, F,
    \prod}$ and $R\subseteq Q \times \mathcal{D}^{\vec{x}} \times Q$ a
  data simulation for $A$, for any tuple $(q,\nu,q') \in R$ we have
  $\mathcal{L}_{(q,\nu)}(A) \subseteq \mathcal{L}_{(q',\nu)}(A)$. 
\end{lemma}
\proof{ Let $(q,\nu)=(q_0,\nu_0) \arrow{\sigma_0}{} \ldots
  \arrow{\sigma_{n-1}}{} (q_n,\nu_n)$ be a run of $A$ such that $q_n
  \in F$. By induction on $n\geq0$ we find a run
  $(q',\nu)=(q'_0,\nu_0) \arrow{\sigma_0}{} \ldots
  \arrow{\sigma_{n-1}}{} (q'_n,\nu_n)$ of $A$, such that
  $(q_i,\nu_i,q'_i) \in R$, for all $i\in[0,n]$. Moreover, since
  $q_n\in F$, we also obtain $q'_n$. \qed}

Let $A = \tuple{\Sigma, \mathcal{D}, \vec{x}, Q, \iota, F, \prod}$,
where $Q = \set{q_1,\ldots,q_k}$, for some $k>0$, be a DA for the rest
of this section. The data simulation algorithm (Algorithm
\ref{alg:data-simulation}) given in this section manipulates sets of
valuations from $\mathcal{D}^{\vec{x}}$ that are definable by
first-order formulae in $\thd$. A relation $R \subseteq Q \times
\mathcal{D}^{\vec{x}} \times Q$ is said to be \emph{definable} if and
only if there exists a matrix $\Phi = [\phi_{ij}]_{i,j=1}^k$ of
formulae $\phi_{ij}(\vec{x}) \in \thd$ such that $(q_i,\nu,q_j) \in R
\iff \nu \models \phi_{ij}$. For $\ell \in [1,k]$, we denote by
$\Phi_\ell$ the $\ell$-th row of the matrix $\Phi$.

Algorithm \ref{alg:data-simulation} is a refinement algorithm which
handles two matrices of formulae that define the relations
$\mathit{Sim}, \mathit{PrevSim} \subseteq Q \times
\mathcal{D}^{\vec{x}} \times Q$. In the following we shall use the
same names to denote the relations and their matrix representations.
Intuitively, $\mathit{PrevSim}$ is the previous candidate for
simulation, whereas $\mathit{Sim}$ is a entry-wise stronger relation,
that refines $\mathit{PrevSim}$. The refinement step is performed
backwards, with respect to each transition rule $q_i
\arrow{\sigma,\phi}{} q_\ell$ of the automaton. Namely, for each pair
of valuations such that $(\nu,\nu') \modelsthd \phi$ and
$(q_\ell,\nu',q_m) \in \mathit{PrevSim}$ for some state $q_m$, we add
the tuple $(q_i,\nu,q_j) \in \mathit{Sim}$ for all predecessors $q_j$
of $q_m$, such that $q_j \arrow{\sigma,\psi}{} q_m$ and $(\nu,\nu')
\modelsthd \psi$. This update guarantees that, for every transition
$(q_i,\nu) \arrow{\sigma}{}_A (q_\ell,\nu')$, where $(q_i,\nu,q_j) \in
\mathit{Sim}$ there exists a state $q_m$ such that $(q_j,\nu)
\arrow{\sigma}{}_A (q_m,\nu')$ and $(q_\ell,\nu',q_m) \in
\mathit{PrevSim}$. The algorithm stops when $\mathit{Sim}$ and
$\mathit{PrevSim}$ define the same relation, which is, moreover, a
simulation.

To define the update, we use the following
function: \[\mathit{PreSim}_\sigma(i,j,\ell,R) \equiv \forall \vec{x}'
~.~ \phi(\vec{x},\vec{x}') \rightarrow \bigvee_{
  \scriptscriptstyle{q_j \arrow{\sigma,\psi}{} q_m}
}\psi(\vec{x},\vec{x}') \wedge R_{\ell m}(\vec{x}'), \text{ where } q
\arrow{\sigma,\phi}{} q' \in \prod\enspace.\] We define also the sets
$\mathit{post}_\sigma(q) = \set{q' \mid q \arrow{\sigma,\phi}{} q' \in
  \prod}$ and $\mathit{pre}_\sigma(q) = \set{q' \mid q'
  \arrow{\sigma,\phi}{} q \in \prod}$. With this notation, Algorithm
\ref{alg:data-simulation} describes the procedure that computes a data
simulation for a given data automaton.

\begin{algorithm}[htb]
\begin{algorithmic}[0]
\State {\bf input}: a data automaton $A =
\tuple{\Sigma,\mathcal{D},\vec{x},Q,\iota,F,\prod}$, where $Q =
\set{q_1,\ldots,q_k}$ and a constant $K>0$

\State {\bf output}: a data simulation $R \subseteq Q \times
\mathcal{D}^{\vec{x}} \times Q$ for $A$ 

\State {\bf global vars} $[\mathit{Sim}_{ij}]_{i,j=1}^k$,
$[\mathit{PrevSim}_{ij}]_{i,j=1}^k$, $[\mathit{Cnt}_{ij}]_{i,j=1}^k$
\end{algorithmic}
\begin{algorithmic}[1]
\For{$i = 1, \ldots, k$}
\For{$j = 1, \ldots, k$}

\State $\mathit{PrevSim}_{ij} \leftarrow \true$\label{ln:initPrevSim}

\State $\mathit{Cnt}_{ij} \leftarrow K$\label{ln:initCnt}

\EndFor

\For{$j= 1, \ldots, k$}

\EndFor

\If{$q_i \in F \text{ and } q_j \not\in F$}

\State $\mathit{Sim}_{ij} \leftarrow \false$\label{ln:simInitFalse}

\Else

\State $\mathit{Sim}_{ij} \leftarrow
\bigwedge_{\sigma\in\Sigma}\bigwedge_{q_\ell\in\mathit{post}_\sigma(q_i)}
\mathit{PreSim}_\sigma(i,j,\ell,\mathit{PrevSim})$\label{ln:simInit}

\EndIf

\EndFor

\For{all $\ell \in [1,k]$ such that $\mathit{Sim}_\ell \not\iff \mathit{PrevSim}_\ell$}\label{ln:simBeginWhile}

\For{$\sigma \in \Sigma$}

\For{$q_i \in \mathit{pre}_\sigma(q_\ell)$}

\For{$j = 1,\ldots,k$}

\State $\mathit{Sim}_{ij} \leftarrow \mathit{Sim}_{ij} \wedge \mathit{PreSim}_\sigma(i,j,\ell,\mathit{Sim})$\label{ln:simUpdate}

\EndFor

\EndFor

\EndFor

\For{all $j=1,\ldots,k$ such that $\mathit{Sim}_{\ell j} \not\iff \mathit{PrevSim}_{\ell j}$}

\If{$\mathit{Cnt}_{\ell j} = 0$}

\State $\mathit{Sim}_{\ell j} \leftarrow \false$\label{ln:simFalse}

\Else

\State $\mathit{Cnt}_{\ell j} \leftarrow \mathit{Cnt}_{\ell j} - 1$\label{ln:cntDec}

\EndIf

\EndFor

\State $\mathit{PrevSim}_\ell \leftarrow \mathit{Sim}_\ell$\label{ln:prevSimUpdate}

\EndFor\label{ln:simEndWhile}

\State {\bf return} $\mathit{Sim}$
\end{algorithmic}
\caption{Data Simulation Algorithm}\label{alg:data-simulation}
\end{algorithm}

Initially, the matrix $\mathit{PrevSim}$ is true everywhere (line
\ref{ln:initPrevSim}). The current simulation candidate $\mathit{Sim}$
is initialized to false for all $i,j \in [1,k]$ such that $q_i \in F$
and $q_j \not\in F$ (line \ref{ln:simInitFalse}). Observe that, in
this case $q_j$ cannot simulate $q_i$, by Definition
\ref{def:data-simulation} (\ref{it1:data-simulation}). Otherwise, we
initialize $\mathit{Sim}_{ij}$ to the strongest pre-simulation with
respect to $\mathit{PrevSim}$ (line \ref{ln:simInit}). In the
iterative loop (lines \ref{ln:simBeginWhile}--\ref{ln:simEndWhile})
the algorithm choses a state $q_\ell$ for which the current simulation
candidate $\mathit{Sim}_\ell$ is not equivalent to the previous one
$\mathit{PrevSim}_\ell$ (line \ref{ln:simBeginWhile}) and sharpens the
set $\mathit{Sim}_{ij}$, with respect to the transition rule $q_i
\arrow{\sigma,\phi}{} q_\ell$, for all input symbols $\sigma\in\Sigma$
and all peer states $q_j,~ j \in [1,k]$ (line \ref{ln:simUpdate}).
The following invariants are key to proving the correctness of
Algorithm \ref{alg:data-simulation}.

\begin{lemma}\label{lemma:sim-invariants}
  The following invariants hold each time Algorithm
  \ref{alg:data-simulation} reaches line \ref{ln:simBeginWhile}:
  \begin{compactitem}
  \item ($\mathit{SimInv}_1$) $\forall i,j \in [1,k] \colon
    \mathit{Sim}_{ij} \rightarrow \mathit{PrevSim}_{ij}\enspace.$
  \item ($\mathit{SimInv}_2$) $\forall \sigma\in\Sigma~ \forall i,j,\ell
    \in [1,k]~ \forall \nu,\nu' \in \mathcal{D}^{\vec{x}} \colon \nu
    \modelsthd \mathit{Sim}_{ij} \text{ and } (q_i,\nu)
    \arrow{\sigma}{} (q_\ell,\nu')~ \Rightarrow \exists m\in[1,k] \colon (q_j,\nu)
    \arrow{\sigma}{} (q_m,\nu') \text{ and } \nu' \modelsthd
    \mathit{PrevSim}_{\ell m}\enspace.$
  \end{compactitem}
\end{lemma}
\proof{ Let $\mathit{Sim}'$ and $\mathit{PrevSim}'$ denote the global
  matrices after one iteration of the loop on lines
  \ref{ln:simBeginWhile}--\ref{ln:simEndWhile}.  

  \vspace*{\baselineskip}\noindent ($\mathit{SimInv}_1$) When line
  \ref{ln:simBeginWhile} is reached for the first time,
  $\mathit{PrevSim}_{ij} = \true$, for all $i,j\in[1,k]$, thus
  $\mathit{SimInv}_1$ holds initially. Since $\mathit{Sim}$ is modified
  only at lines \ref{ln:simUpdate} or \ref{ln:simFalse}, we have
  $\mathit{Sim}_{ij} \rightarrow \mathit{Sim}'_{ij}$, for all
  $i,j\in[1,k]$. Moreover, either $\mathit{PrevSim}'_{ij} =
  \mathit{Sim}_{ij}$, or $\mathit{PrevSim}'_{ij} =
  \mathit{PrevSim}_{ij}$, for all $i,j \in [1,k]$ (line
  \ref{ln:prevSimUpdate}). Thus $\mathit{PrevSim}'_{ij} \rightarrow
  \mathit{Sim}_{ij} \rightarrow \mathit{Sim}'_{ij}$, for all
  $i,j\in[1,k]$, by an application of the induction hypothesis.

  \vspace*{\baselineskip}\noindent ($\mathit{SimInv}_2$) We show
  that this invariant holds the first time the control reaches line
  \ref{ln:simBeginWhile}. Let $\sigma\in\Sigma$, $i,j,\ell\in[1,k]$
  and $\nu,\nu' \in \mathcal{D}^{\vec{x}}$ such that $\nu \modelsthd
  \mathit{Sim}_{ij}$ and $(q_i,\nu) \arrow{\sigma}{}
  (q_\ell,\nu')$. Since $\nu \modelsthd \mathit{Sim}_{ij}$ (thus
  $\mathit{Sim}_{ij}\neq\false$) and
  $q_\ell\in\mathit{post}_\sigma(q_i)$ we have that $\nu \modelsthd
  \mathit{PreSim}_\sigma(i,j,\ell,\mathit{PrevSim})$, where $q_i
  \arrow{\sigma,\phi}{} q_\ell \in \Delta$. Since $(q_i,\nu)
  \arrow{\sigma}{} (q_\ell,\nu')$ we obtain that $(\nu,\nu')
  \modelsthd \phi(\vec{x},\vec{x}')$, and consequently $(\nu,\nu')
  \modelsthd \psi(\vec{x},\vec{x}') \wedge \mathit{PrevSim}_{\ell
    m}(\vec{x}')$, for some $m \in [1,k]$, such that $q_j
  \arrow{\sigma,\psi}{} q_m \in \prod$. Hence $\mathit{SimInv}_2$
  holds when the control first reaches line \ref{ln:simBeginWhile}.

  For the induction step, let us assume that $\mathit{SimInv}_2$ holds at
  line \ref{ln:simBeginWhile} and we prove that it holds also after
  executing line \ref{ln:prevSimUpdate}. Let $\sigma\in\Sigma$,
  $i,j,\ell\in[1,k]$ and $\nu,\nu' \in \mathcal{D}^{\vec{x}}$ such
  that $\nu \modelsthd \mathit{Sim}'_{ij}$ and $(q_i,\nu)
  \arrow{\sigma}{} (q_\ell,\nu')$. We distinguish two cases:
  \begin{compactenum}
  \item if $\mathit{Sim}_\ell \not\iff \mathit{PrevSim}_\ell$ on line
    \ref{ln:simBeginWhile}, since $q_i \in
    \mathit{pre}_\sigma(q_\ell)$, then $\mathit{Sim}'_{ij}$ was
    updated at line \ref{ln:simUpdate}. Since $\nu \modelsthd
    \mathit{Sim}'_{ij}$, we obtain $\nu \modelsthd
    \mathit{PreSim}_\sigma(q_i,q_j,q_\ell,\mathit{Sim})$. Moreover,
    $\mathit{PrevSim}'_\ell$ is updated to $\mathit{Sim}'_\ell$ at
    line \ref{ln:prevSimUpdate}, hence $\nu \modelsthd
    \mathit{PreSim}_\sigma(q_i,q_j,q_\ell,\mathit{PrevSim}')$ as well.
    Since $(q_i,\nu) \arrow{\sigma}{} (q_\ell,\nu')$, we obtain that
    $(\nu,\nu') \modelsthd \psi(\vec{x},\vec{x}') \wedge
    \mathit{PrevSim}'_{\ell m}(\vec{x}')$, for some $m\in[1,k]$ such
    that $q_j \arrow{\sigma,\psi}{} q_m \in \prod$, thus $(\nu,\nu')
    \modelsthd \psi(\vec{x},\vec{x}')$ and $\nu' \modelsthd
    \mathit{PrevSim}'_{\ell m}$. Thus $\mathit{SimInv}_2$ holds
    for $\mathit{Sim}'$ and $\mathit{PrevSim}'$.
  \item else $\mathit{Sim}_\ell \iff \mathit{PrevSim}_\ell$ on line
    \ref{ln:simBeginWhile}, $\mathit{PrevSim}'_\ell \iff
    \mathit{PrevSim}_\ell$ because the update on line
    \ref{ln:prevSimUpdate} is skipped, and for all $q_i \in
    \mathit{pre}_\sigma(q_\ell)$ and all $j \in [1,k]$, we have
    $\mathit{Sim}'_{ij} \iff \mathit{Sim}_{ij}$. Then $\mathit{SimInv}_2$
    holds for $\mathit{Sim}'$ and $\mathit{PrevSim}'$ because it holds
    for $\mathit{Sim}$ and $\mathit{PrevSim}$, by the induction
    hypothesis. 
  \end{compactenum}
\qed}

The algorithm iterates the loop on lines
(\ref{ln:simBeginWhile}--\ref{ln:simEndWhile}) until $\mathit{Sim}$
and $\mathit{PrevSim}$ define the same relation. Since, in general the
data constraints $\mathit{Sim}_{ij}$, at each iteration step, might
form an infinitely decreasing sequence, we use the matrix
$\mathit{Cnt}$ of integer counters, initially set to some input value
$K>0$ (line \ref{ln:initCnt}). Observe that each entry
$\mathit{Cnt}_{ij}$ decreases every time $\mathit{Sim}_{ij} \not\iff
\mathit{PrevSim}_{ij}$ (line \ref{ln:cntDec}). When the counter
$\mathit{Cnt}_{ij}$ reaches zero, we set $\mathit{Sim}_{ij}$ to false
(line \ref{ln:simFalse}), which guarantees that $\mathit{Sim}_{ij}
\iff \mathit{PrevSim}_{ij}$ always in the future. Since the number of
entries in the counter matrix is finite, the algorithm is guaranteed
to terminate. The following theorem summarizes the main result of this
section.

\begin{theorem}\label{thm:simulation}
  Let $A = \tuple{\Sigma,\mathcal{D},\vec{x},Q,\iota,F,\prod}$ be a
  DA. Then Algorithm \ref{alg:data-simulation} terminates on input $A$
  and the output is a data simulation $R \subseteq Q \times
  \mathcal{D}^{\vec{x}} \times Q$ for $A$. 
\end{theorem}
\proof{ Let $\mathit{Sim}^n$ and $\mathit{PrevSim}^n$ denote the
  matrices $\mathit{Sim}$ and $\mathit{PrevSim}$ at the $n$-th
  iteration of the loop on lines
  \ref{ln:simBeginWhile}--\ref{ln:simEndWhile}, for
  $n\geq0$. Algorithm \ref{alg:data-simulation} terminates whenever
  $\mathit{Sim}^n_{ij} \iff \mathit{PrevSim}^n_{ij}$, for all
  $i,j\in[1,k]$ (line \ref{ln:simBeginWhile}). Suppose, by
  contradiction, that this never happens, thus there exist
  $i,j\in[1,k]$ such that $\mathit{Sim}^n_{ij} \not\iff
  \mathit{PrevSim}^n_{ij}$, for all $n\geq0$. Then
  $\mathit{Cnt}_{ij}^K = 0$ (line \ref{ln:cntDec}) and
  $\mathit{Sim}_{ij}^{K+1} = \mathit{PrevSim}_{ij}^{K+2} = \false$
  (lines \ref{ln:simFalse} and \ref{ln:prevSimUpdate}). Since
  $\mathit{Sim}^n_{ij} \rightarrow \mathit{PrevSim}^n_{ij}$, by Lemma
  \ref{lemma:sim-invariants} ($\mathit{SimInv}_1$), we obtain that
  $\mathit{Sim}^{K+2}_{ij} = \mathit{PrevSim}^{K+2}_{ij}$,
  contradiction. 

  To prove that the output of Algorithm \ref{alg:data-simulation} is a
  data simulation for $A$, we use Lemma \ref{lemma:sim-invariants}
  ($\mathit{SimInv}_2$) and the fact that, upon termination, we
  have $\mathit{Sim}_{ij} \iff \mathit{PrevSim}_{ij}$, for all
  $i,j\in[1,k]$. \qed}

\subsection{Simulation and Subsumption}

Finally, we explain how a data simulation relation computed by
Algorithm \ref{alg:data-simulation} can be used to optimize the trace
inclusion semi-algorithm. Let $\mathcal{A} = \tuple{A_1, \ldots, A_N}$
be DAN, where $A_i =
\tuple{\mathcal{D},\Sigma_i,\vec{x}_i,Q_i,\iota_i,F_i,\prod_i}$, for
all $i \in [1,N]$, and $B =
\tuple{\mathcal{D},\Sigma,\vec{x}_B,Q_B,\iota_B,F_B,\prod_B}$ be an
observer DA such that $\vec{x}_B \subseteq \bigcup_{i=1}^N\vec{x}_i$.

The main problem in using data simulation to enhance the convergence
of our trace inclusion semi-algorithm is related to the fact that
simulation relations are, in general, not compositional w.r.t.  the
interleaving semantics of the network. In other words, if we have $N$
data simulations $R_i \subseteq Q_i \times \mathcal{D}^{\vec{x}_i}
\times Q_i$, then their cross-product $R \subseteq Q_{\mathcal{A}}
\times \mathcal{D}^{\vec{x}_{\mathcal{A}}} \times Q_{\mathcal{A}}$
defined as: \[\forall q_1,r_1\in Q_1 \ldots \forall q_N,r_N \in Q_N
\forall \nu \in \mathcal{D}^{\vec{x}_{\mathcal{A}}} \colon
(\tuple{q_1,\ldots,q_N},\nu,\tuple{r_1,\ldots,r_N}) \in R \iff
(q_i,\nu\proj_{\vec{x}_i},r_i) \in R_i\] is not necessarily a
simulation on the network expansion $\mathcal{A}^e$. The reason for
this can be seen for $N=2$. Let $\sigma_1,\sigma_2 \in
\Sigma_{\mathcal{A}}$, such that $\sigma_1 \not\in \Sigma_2$ and
$\sigma_2 \not\in \Sigma_1$. The execution of $\mathcal{A}^e$ on the
sequence of input symbols $\sigma_1\sigma_2$ is $(\tuple{q_1,q_2},\nu)
\arrow{\sigma_1}{} (\tuple{q'_1,q_2},\nu') \arrow{\sigma_2}{}
(\tuple{q'_1,q'_2},\nu'')$. Suppose that
$(q_i,\nu\proj_{\vec{x}_i},r_i) \in R_i$, $i=1,2$. Then there exists
$r'_1 \in Q_1$ such that $(\tuple{r_1,r_2},\nu) \arrow{\sigma_1}{}
(\tuple{r'_1,r_2},\nu')$ and $(q'_1,\nu'\proj_{\vec{x}_1},r'_1) \in
R_1$. In order to use the simulation and build the continuation
$(\tuple{r'_1,r_2},\nu') \arrow{\sigma_2}{}
(\tuple{r'_1,r'_2},\nu'')$, we would need that
$(q_2,\nu'\proj_{\vec{x}_2},r_2) \in R_2$, which is not necessarily
ensured by the hypothesis $(q_2,\nu\proj_{\vec{x}_2},r_2) \in R_2$.

We propose a partial solution to this problem, based on a restriction
concerning the distribution of the network variables
$\vec{x}_{\mathcal{A}} = \bigcup_{i=1}^N \vec{x}_i$ over the
components $A_1,\ldots,A_N$: for each $i\in[1,N]$, we have $\vec{x}_i
= \vec{x}^g \cup \vec{x}^\ell_i$, where $\vec{x}^g$ is a set of
\emph{global} variables, and $\vec{x}^\ell_i$ are the \emph{local}
variables of $A_i$. In other words, the only variables shared between
more than one component are $\vec{x}^g$, which, moreover, are visible
to all components.

Then the problem can be bypassed if none of the simulation relations
$R_i \subseteq Q_i \times \mathcal{D}^{\vec{x}_i} \times Q_i$
may constrain the global variables:

\begin{ass}\label{ass:global-unrestricted}
For each $i\in[1,N]$ and each $(q_i,\nu,r_i) \in R_i$ we also have
$(q_i,\nu',r_i) \in R_i$ for each $\nu'\in\mathcal{D}^{\vec{x}_i}$
such that $\nu\proj_{\vec{x}^\ell_i}=\nu'\proj_{\vec{x}^\ell_i}$.
\end{ass}

Under this assumption, we use pre-computed data simulations $R_i
\subseteq Q_i \times \mathcal{D}^{\vec{x}^g} \times Q_i$ and $R_B
\subseteq Q_B \times \mathcal{D}^{\vec{x}_B} \times Q_B$ to generalize
the basic subsumption relation between product states (defined by
Lemma \ref{lemma:img-subsumption}) thus speeding up the convergence of
Algorithm \ref{alg:trace-inclusion}.

\begin{lemma}\label{lemma:sim-subsumption}
  Under assumption \ref{ass:global-unrestricted}, the relation defined
  as: \[\begin{array}{c}
  (\tuple{q_1,\ldots,q_N},P,\Phi) \sqsubseteq_{\mathit{sim}}
  (\tuple{r_1,\ldots,r_N},S,\Psi) \\
  \iff \\
  \forall i \in [1,N] ~\forall \nu \in \mathcal{D}^{\vec{x}_{\mathcal{A}}}\colon
  \nu \models \Phi \Rightarrow \nu \models \Psi \text{ and }
  \left\{\begin{array}{rcl}
  (q_i,\nu\proj_{\vec{x}_i},r_i) & \in & R_i \\
  \forall p \in S \exists q \in P ~.~ (p,\nu\proj_{\vec{x}_B},q) & \in & R_B
  \end{array}\right. 
  \end{array}\] 
  is a subsumption relation. 
\end{lemma}
\proof{ Let $s=(\tuple{q_1,\ldots,q_N},P,\Phi)$ and
  $t=(\tuple{r_1,\ldots,r_N},S,\Psi)$ be two product states, such that
  $s \sqsubseteq_{\mathit{sim}} t$. According to Definition
  \ref{def:subsumption}, we need to prove that
  $\mathcal{L}_s(\mathcal{A}^e \times \overline{B}) \subseteq
  \mathcal{L}_t(\mathcal{A}^e \times \overline{B})$. To this end, it
  is sufficient to prove that for each $\nu \in
  \mathcal{D}^{\vec{x}_{\mathcal{A}}}$ such that $\nu \models \Phi$:
  \begin{compactenum}
    \item $\mathcal{L}_{(\tuple{q_1,\ldots,q_N},\nu)}(\mathcal{A}^e)
      \subseteq
      \mathcal{L}_{(\tuple{r_1,\ldots,r_N},\nu)}(\mathcal{A}^e)$, and
    \item for all $p \in S$ there exists $q \in P$ such that
      $\mathcal{L}_(p,\nu\proj_{\vec{x}_B})(B) \subseteq
      \mathcal{L}_(q,\nu\proj_{\vec{x}_B})(B)$. 
  \end{compactenum}
  Assuming that the above statements hold, we have:
  \[\begin{array}{rcl} \mathcal{L}_s(\mathcal{A}^e \times \overline{B}) & = & 
  \bigcup_{\nu \models \Phi} \left(\mathcal{L}_{(\tuple{q_1,\ldots,q_N},\nu)}(\mathcal{A}^e) \cap 
  \bigcap_{q \in P} \mathcal{L}_{(q,\nu\proj_{\vec{x}_B})}(\overline{B})\right) \\
  & \subseteq & 
  \bigcup_{\nu \models \Phi} \left(\mathcal{L}_{(\tuple{r_1,\ldots,r_N},\nu)}(\mathcal{A}^e) \cap 
  \bigcap_{p \in S} \mathcal{L}_{(p,\nu\proj_{\vec{x}_B})}(\overline{B})\right) \\
  & \subseteq & 
  \bigcup_{\nu \models \Psi} \left(\mathcal{L}_{(\tuple{r_1,\ldots,r_N},\nu)}(\mathcal{A}^e) \cap 
  \bigcap_{p \in S} \mathcal{L}_{(p,\nu\proj_{\vec{x}_B})}(\overline{B})\right) \\
  & = & \mathcal{L}_t(\mathcal{A}^e
   \times \overline{B})
   \end{array}\] and we are done. Moreover, the second point
   above is a direct consequence of the second point of the definition
   of $\sqsubseteq_{\mathit{sim}}$ and Definition
   \ref{def:subsumption}. We are left with proving the first
   point. Let $(\tuple{q_1,\ldots,q_N},\nu) \arrow{\sigma}{}
   (\tuple{q'_1,\ldots,q'_N},\nu')$ be a transition of $\mathcal{A}^e$
   and let $(\tuple{r_1,\ldots,r_N}, \nu)$ be a configuration of
   $\mathcal{A}^e$ such that $(q_i,\nu\proj_{\vec{x}_i},r_i) \in R_i$,
   for each $i\in[1,N]$. Let $i \in [1,N]$ be an arbitrary component,
   and distinguish two cases: \begin{compactitem}
   \item if $q_i \arrow{\sigma,\phi_i}{} q'_i \in \prod_i$ and
     $(\nu\proj_{\vec{x}_i},\nu'\proj_{\vec{x}_i}) \modelsthd \phi_i$,
     i.e.\ $(q_i,\nu\proj_{\vec{x}_i}) \arrow{\sigma}{}
     (q'_i,\nu'\proj_{\vec{x}_i})$, then, since
     $(q_i,\nu\proj_{\vec{x}_i},r_i) \in R_i$ there exists $r'_i \in
     Q_i$ s.t.\ $(r_i,\nu\proj_{\vec{x}_i}) \arrow{\sigma}{}
     (r'_i,\nu\proj_{\vec{x}_i})$ and
     $(q'_i,\nu\proj_{\vec{x}_i},r'_i) \in R_i$.
   \item otherwise, $q_i = q'_i$ and
     $\nu\proj_{\vec{x}^\ell_i}=\nu'\proj_{\vec{x}^\ell_i}$. By
     Assumption \ref{ass:global-unrestricted}, we obtain
     $(q_i,\nu'\proj_{\vec{x}_i}, r_i) \in R_i$. By chosing
     $r'_i=r_i$, we obtain $(q'_i,\nu'\proj_{\vec{x}_i}, r'_i) \in R_i$.
   \end{compactitem}
   Hence $(q'_i,\nu'\proj_{\vec{x}_i},r'_i) \in R_i$, for all $i \in
   [1,N]$. Thus, the relation defined as:
   \[(\tuple{q_1,\ldots,q_N},\nu,\tuple{r_1,\ldots,r_N}) \in R 
   \iff \forall i\in[1,N]\colon (q_i,\nu\proj_{\vec{x}_i},r_i) \in
   R_i\] is a data simulation (Definition \ref{def:data-simulation}),
   thus, by Lemma \ref{lemma:simulation-residual}, we obtain that
   $\mathcal{L}_{(\tuple{q_1,\ldots,q_N},\nu)}(\mathcal{A}^e)
   \subseteq
   \mathcal{L}_{(\tuple{r_1,\ldots,r_N},\nu)}(\mathcal{A}^e)$, for all
   $\nu\in\mathcal{D}^{\vec{x}_{\mathcal{A}}}$, such that $\nu
   \modelsthd \Phi$, and the first point above holds. \qed}

\section{Experimental Results}\label{sec:experiments}

We have implemented Algorithm \ref{alg:trace-inclusion} in a prototype
tool\footnote{{\tt http://www.fit.vutbr.cz/research/groups/verifit/tools/includer/}}
using the \textsc{MathSat} SMT solver \cite{mathsat} for answering the
satisfiability queries and computing the interpolants. The results of
the experiments are given in Tables~\ref{TaExpRes} and
~\ref{TaExpResProd}. The results were obtained on an Intel i7-4770 CPU
@ 3.40GHz machine with 32GB RAM.

\begin{table}
\begin{center}
{\fontsize{8}{9}\selectfont
  \caption{Experiments with single-component networks.}
  \label{TaExpRes}
\begin{tabular}{||l|c|c|c|c||c|c||}
  \hline
  Example & $A$ ($|Q|$/$|\Delta|$) & $B$ ($|Q|$/$|\Delta|$) & Vars. & Res.  & Time \\
  \hline \hline
Arrays shift        & 3/3  & 3/4 & 5 &ok           &  $<0.1s$ \\
  \hline
Array rotation 1 &  4/5  & 4/5 & 7 &ok    & $0.1s$ \\ 
  \hline
Array rotation 2 &  8/21 & 6/24& 11 &ok    &  $34s$ \\ 
  \hline
Array split      &  20/103 & 6/26& 14 &ok    &  $4m32s$ \\ 
  \hline
HW counter 1      &  2/3 & 1/2 & 2 &ok           & $0.2s$  \\
  \hline
HW counter 2            &  6/12 & 1/2 & 2 &ok           & $0.4s$  \\
  \hline
Synchr. LIFO       &  4/34 & 2/15 & 4 &ok           &  $2.5s$ \\
  \hline
ABP-error              & 14/20 & 2/6& 14 &cex  &  $2s$   \\
  \hline
ABP-correct          &  14/20 & 2/6& 14&ok           &  $3s$ \\
  \hline
\end{tabular}
}
\end{center}
\end{table}

Table ~\ref{TaExpRes} contains experiments where the network
$\mathcal{A}$ consists of a single component. We applied the tool on
several verification conditions generated from imperative programs
with arrays \cite{cav09} (Array shift, Array rotation 1+2, Array
split) available online \cite{ntslib}. Then, we applied it on models
of hardware circuits (HW Counter 1+2, Synchronous LIFO)
\cite{smrcka}. Finally, we checked two versions (correct and faulty)
of the timed Alternating Bit Protocol \cite{abp}.

Table ~\ref{TaExpResProd} provides a list of experiments where the
network $\mathcal{A}$ has $N>1$ components. First, we have the example
of Fig. \ref{fig:running-example} (Running). Next, we have several
examples of real-time verification problems \cite{stavros-thesis}:
a~controller of a railroad crossing \cite{henzinger:RealTimeSystems}
(Train) with $T$ trains, the Fischer Mutual Exclusion protocol with
deadlines $\Delta$ and $\Gamma$ (Fischer), and a~hardware
communication circuit with $K$ stages, composed of timed NOR gates
(Stari). Third, we have modelled a Producer-Consumer example
\cite{AmitThesis} with a fixed buffer size $B$. Fourth, we have
experimented with several models of parallel programs that manipulate
arrays (Array init, Array copy, Array join) with window size
$\Delta$.

\begin{table}[t]
\begin{center}
{\fontsize{8}{9}\selectfont
  \caption{Experiments with multiple-component networks (e.g.,
    $2\times 2/2 + 2\times 3/3$ in column $\mathcal{A}$ means that
    $\mathcal{A}$ is a~network with $4$ components, of which $2$ DA with 2 states
    and $2$ rules, and $2$ DA with $3$ states and $3$ rules).}
\label{TaExpResProd}
\begin{tabular}{||l|c|c|c|c|c|c||c|c||}
  \hline
  Example & N &$\mathcal{A}$ ($|Q|$/$|\Delta|$) & $B$ ($|Q|$/$|\Delta|$) & Vars. & Res.  & Time \\
  \hline \hline
Running & 2 & 2$\times$2/2  & 3/4& 3 &ok & $0.2s$\\
\hline
Running & 10 & 10$\times$2/2  & 11/20& 3 &ok & $25s$\\
\hline
Train ($T=5$) & 7 & 5$\times$3/3 + 4/4 + 4/4  & 2/38& 1 &ok & $4s$\\
\hline
Train ($T=20$) & 22 & 20$\times$3/3 + 4/4 + 4/4  & 2/128& 1 &ok & $6m26s$\\
\hline
Fischer ($\Delta=1$, $\Gamma=2$)  & 2 & 2$\times$5/6  & 1/10& 4 &ok & $8s$\\
\hline
Fischer ($\Delta=1$, $\Gamma=2$)  & 3 & 3$\times$5/6  & 1/15& 4 &ok & $2m48s$\\
\hline
Fischer ($\Delta=2$, $\Gamma=1$)  & 2 & 2$\times$5/6  & 1/10& 4 &cex & $3s$\\
\hline
Fischer ($\Delta=2$, $\Gamma=1$)  & 3 & 3$\times$5/6  & 1/15& 4 &cex & $32s$\\
\hline
Stari ($K=1$)  & 5 &  4/5 + 2/4 + 5/7 + 5/7 + 5/7  & 3/6 & 3&ok & $0.5s$\\
\hline
Stari ($K=2$)  & 8 &  4/5 + 2/4 + 2$\times$5/7 + 2$\times$5/7 + 2$\times$5/7  & 3/6 & 3&ok & $0.5s$\\
\hline
Prod-Cons ($B=3$)  & 2 & 4/4 + 4/4  & 2/7 & 2 &ok & $10s$\\
\hline
Prod-Cons ($B=6$)  & 2 & 4/4 + 4/4  & 2/7 & 2 &ok & $2m32s$\\
\hline
Array init ($\Delta=2$) & 5 & 5$\times$2/2  & 2/6 & 2 &ok & $3s$\\
\hline
Array init ($\Delta=2$) & 15 & 15$\times$2/2  & 2/16 & 2 &ok & $3m15s$\\
\hline
Array copy ($\Delta=20$) & 20 & 20$\times$2/2  & 2/21 & 3 &ok & $0.3s$\\
\hline
Array copy ($\Delta=20$) & 150 & 150$\times$2/2  & 2/151 & 3 &ok & $43s$\\
\hline
Array join ($\Delta=10$)  & 4 & 2$\times$2/2 + 2$\times$3/3  & 2/3 & 2 &ok & $6s$\\
\hline
Array join ($\Delta=20$)  & 6 & 3$\times$2/2 + 3$\times$3/3  & 2/4 & 2 &ok & $1m9s$\\
\hline

\end{tabular}
}
\end{center}
\end{table}

For the time being, our implementation is a proof-of-concept prototype
that leaves plenty of room for optimization (e.g.\ caching
intermediate computation results) likely to improve the performance on
more complicated examples. Despite that, we found the results from
Tables~\ref{TaExpRes} and \ref{TaExpResProd} rather encouraging.

\section{Conclusions}

We have presented an interpolation-based abstraction refinement method
for trace inclusion between a network of data automata and an
observer where the variables used by the observer are a subset of
those used by the network. The procedure builds on a new
determinization result for DAs and combines in a novel way predicate
abstraction and interpolation with antichain-based inclusion checking.
The procedure has been successfully applied to several examples,
including verification problems for array programs, real-time systems,
and hardware designs. Future work includes an extension of the method
to data tree automata and its application to logics for heaps with
data. Also, we foresee an extension of the method to handle infinite
traces.

\paragraph{Acknowledgement.} This work was supported by the Czech Science
Foundation project 14-11384S, the BUT FIT project FIT-S-14-2486, and the French
ANR-14-CE28-0018 project Vecolib.

\bibliographystyle{splncs03}
\bibliography{ref}

\end{document}